\documentclass[fleqn,10pt]{wlscirep}
\usepackage[utf8]{inputenc}
\usepackage{bm}
\usepackage{lineno}

\title{The Higgs boson - its implications and prospects for future discoveries}

\author[1,2]{Steven D. Bass}
\author[3,4]{Albert De Roeck}
\author[5,6]{Marumi Kado}
\affil[1]{Kitzb\"uhel Centre for Physics, Kitzb\"uhel, Austria.}
\affil[2]{
Jagiellonian University,
Marian Smoluchowski Institute of Physics,  Krak\'ow, Poland.}
\affil[3]{CERN, Experimental Physics Department, Geneva, Switzerland.}
\affil[4]{University of Antwerp, Physics Department, Antwerp, Belgium.}
\affil[5]{INFN Sezione di Roma and Dipartimento di Fisica, Sapienza Università di Roma, Rome, Italy.}
\affil[6]{IJCLab, Université Paris-Saclay, CNRS/IN2P3, 91405, Orsay, France.}

\begin{abstract}
The Higgs boson, a fundamental scalar, was discovered at CERN in 2012 with mass 125 GeV, a mass that turned out to be a remarkable choice of Nature.
In the Standard Model of particle physics, the Higgs boson is closely linked to the mechanism that gives mass 
to the 
W and Z gauge bosons that mediate the 
weak interactions and to the charged fermions.
Following 
discovery of the Higgs boson, 
present measurements at the Large Hadron Collider are focused on testing the Higgs boson's couplings
to other elementary particles,
precision measurements of the Higgs boson's properties and 
initial 
investigation
of the Higgs boson's self-interaction and shape of the Higgs potential.
With the Higgs boson mass of 125 GeV the 
vacuum sits very close to the border of stable and metastable, which may be a 
hint to deeper physics beyond
the Standard Model.
The Higgs potential 
also plays an important role in 
ideas about 
the cosmological constant or dark energy that drives the accelerating expansion of the Universe, the mysterious dark matter that comprises about 80\% of the matter component in the Universe, as well as a possible phase transition in the early Universe that might be responsible for baryogenesis.
Detailed study of the Higgs boson is at the centre of the recent European Strategy for Particle Physics update.
Here we review the present status of this physics and discuss the new insights expected from present and
future experiments.
\end{abstract}

\begin{document}

\flushbottom
\maketitle

\thispagestyle{empty}

\noindent \textbf{Key points:} 

\noindent
$\bullet$
The discovery of the Higgs boson, the first ever 
observed elementary scalar particle, at CERN's Large Hadron Collider (LHC) in 2012 was a major milestone in the development of particle physics.

\noindent{
$\bullet$
Besides confirming
the mechanism which gives
masses to the W and Z bosons,
thus making the electroweak interaction short range,
direct tests of the couplings of the Higgs boson to fermions are major achievements of the LHC program.
A recent highlight is direct observation of the Higgs boson coupling to muons.
}

\noindent{
$\bullet$
The observed properties of the Higgs boson
put the Standard Model vacuum intriguingly very close to the border of stable and metastable.
Further connections to the outstanding questions of baryogenesis, dark matter, dark energy and inflation mean that the Higgs boson is central to our understanding of the physical Universe.
}

\noindent{
$\bullet$
Precision measurements of the Higgs boson to further probe its 
interactions and possible deeper origin and structure
are an essential part of 
the High-Luminosity LHC program and were recently identified 
by the European Strategy for Particle Physics
to be the highest priority for the next high-energy collider facility.
}

\vspace{3ex}

\section{Introduction} 
\label{sec:intro}

The discovery of the Higgs boson
in 2012 at CERN's Large Hadron Collider, LHC,
in Geneva Switzerland
by the 
ATLAS \cite{Aad:2012tfa} and CMS
\cite{Chatrchyan:2012ufa} experiments
was a key milestone for particle physics,
rewarded by the award of the 2013 Nobel Prize for Physics
to Fran{\c c}ois Englert
\cite{Englert:2014zpa}
and Peter Higgs
\cite{Higgs:2014aqa}.
The Higgs boson
is central to our
understanding of
particle physics.
It is the first
(and so far only) discovered seemingly elementary particle 
with spin zero.

The Standard Model 
\cite{Altarelli:2013tya,Pokorski:1987ed,Aitchison:2004cs}
provides an excellent description of 
particle physics experimental results so far,
from collider experiments at the LHC \cite{Altarelli:2013lla}, with 
centre of mass energy
up to 13 TeV \cite{Foot1},
to low-energy precision measurements
including those of the fine structure constant~\cite{Hanneke:2008tm,Parker:2018vye} 
of Quantum Electrodynamics, QED, 
and the electron's electric dipole moment \cite{Andreev:2018ayy}.
The matter of everyday experience
is built of elementary fermions: quarks and leptons.
Particle
interactions 
are determined by local gauge symmetries
and mediated by the exchange of spin-one gauge bosons.
These are the massless photon for QED
which binds electrons to nuclei in atoms,
gluons for Quantum Chromodynamics, 
QCD, which bind quarks inside the proton and the massive W and Z bosons for the weak interactions
that power the sun and nuclear reactors.
Symmetry drives the particle interactions.
Invariance under local changes in the phases of fermion fields
guarantees the dynamics.
An important ingredient is the origin of particle masses.
Within the Standard Model the masses of 
the W and Z gauge bosons and charged fermions
come from coupling of these particles
to the scalar spin-zero Higgs field which comes with a non-vanishing vacuum expectation value, vev, and a Higgs condensate filling all space.

While the discovered boson behaves very much like 
the Standard Model Higgs with a mass of 125 GeV, 
in which case it completes the
particle spectrum of the Standard Model, 
important open puzzles remain 
connecting 
particle physics to cosmology that require extra new physics.
These are the
dark energy that generates the accelerating expansion of the Universe 
\cite{Frieman:2008sn},
the matter-antimatter asymmetry in the Universe \cite{Dine:2003ax}, 
primordial inflation \cite{Baumann:2008bn}
as well as 
the mysterious extra dark matter which comprises about 80\% of the matter component
in the Universe \cite{Baudis:2018bvr}.
Considerable theoretical work has gone into thinking about possible connections between these open issues and the properties of the Higgs boson.
The Higgs boson's observed decays to vector bosons indicate the existence of a Higgs boson condensate. 
While its mass was expected to be commensurate with the electroweak scale to ensure unitarity of the scattering of longitudinally polarized vector bosons, such a relatively small mass,
which is
very much less than the Planck scale that defines the limit of particle physics 
before quantum gravity effects might apply,
raised the fundamental question of the naturalness of the Standard Model. 

The European Particle Physics  Strategy recently identified precision studies of the Higgs boson as the main priority for the next high-energy collider with measurements
first at the planned high luminosity upgrade of the LHC and, later, with a dedicated Higgs factory as a new facility.
This program involves essential interaction between experiment and theory.
How well does the discovered boson match the Standard Model Higgs and can we find hints for new physics beyond the Standard Model?

This article surveys 
Higgs boson physics with an outlook to future experiments.
In Section \ref{sec:higssth}
we discuss the role of the Higgs boson in the origin of mass.
Section \ref{Discovery} reviews the discovery and early
measurements 
of the Higgs boson's properties.
Recent measurements of the Higgs coupling to fermions are discussed in Section 
\ref{fermions}.
Section \ref{properties} summarises the status of measurement of the Higgs boson's properties and 
interactions
in comparison to the predictions for the Higgs boson
described by the
Standard Model.
In Section \ref{lambda} 
we discuss
the Higgs self-coupling.
Section \ref{extrahiggses} 
focuses on 
searches for any extra Higgs states or possible new CP violation in the Higgs sector.
In Sections \ref{sec:vac} and \ref{sec:cosmo}
we describe open
theoretical issues connected to the Higgs boson in particle physics and cosmology.
What might the Higgs boson be telling us about (the need for) extra new physics beyond the Standard Model?
Finally, in Section \ref{conclusions} we give an outlook to future 
measurements
that might shed light on these questions and the role of the Higgs in understanding the deep structure of the Universe.

\section{The Higgs boson and massive gauge bosons}
\label{sec:higssth}

The Higgs story
starts from the interplay of mass and gauge invariance.
If taken alone,
mass terms for gauge bosons break
the underlying gauge symmetry.
For example, 
consider particles
(fermions or scalar bosons)
${\rm \chi}$ interacting with a gauge field 
${\rm A_{\mu}}$ with
the system 
invariant under
the local gauge transformations
${\rm \chi \to e^{i \alpha} \chi}$
and
${\rm A_{\mu}} \to {\rm A_{\mu}} + \frac{1}{g} 
\partial_{\mu} {\rm \alpha}$
where $g$ is the coupling of ${\rm A_{\mu}}$ to ${\rm \chi}$.
Introducing a
mass term 
$m^2 
{\rm A_{\mu} A^{\mu}}$ 
violates the gauge symmetry without extra ingredients.

This problem is resolved through the Brout-Englert-Higgs (BEH) mechanism
\cite{Higgs:1964ia,Higgs:1964pj,Higgs:1966ev,Englert:1964et}
with related work in Refs\cite{Guralnik:1964eu,Kibble:1967sv}.
The gauge symmetry of the underlying theory can be hidden in the ground state.
The symmetry parameter ${\rm \alpha}$
freezes out to a
particular value 
with
all possible values being degenerate. 
This process,
which is known as
spontaneous symmetry breaking,
generates massless Goldstone modes -- one for each generator of the symmetry. For local gauge symmetries these massless Goldstone modes combine with the gauge bosons to generate new longitudinal modes of the gauge fields, conserving the total number of degrees of freedom. %
The transverse and longitudinal components of the spin-one gauge field acquire non-zero mass, which is the same for
both components. 
In addition, a new scalar boson is produced with finite coupling to the massive gauge fields -- the Higgs boson.

In the Standard Model of particle physics,
besides giving mass to the W and Z gauge bosons, the BEH
mechanism
also plays a vital role with ensuring 
consistent ultraviolet
behaviour of 
scattering amplitudes.
The Higgs boson with mass 125 GeV guarantees unitarity of high energy collisions involving 
massive W and Z bosons, with the Higgs boson
cancelling terms 
from the longitudinal component of the W and Z bosons
that would otherwise violate perturbative unitarity 
~\cite{LlewellynSmith:1973yud,Bell:1973ex,Cornwall:1973tb,Cornwall:1974km}. 
It is also essential
for the renormalizability of the theory, viz.
consistent treatment of ultraviolet divergences 
which 
appear in 
Feynman diagrams
involving loops
\cite{tHooft:1971qjg,tHooft:1972tcz,Veltman:1968ki}.

\begin{figure}[t!]  
\centerline
{\includegraphics[width=0.45\textwidth]
{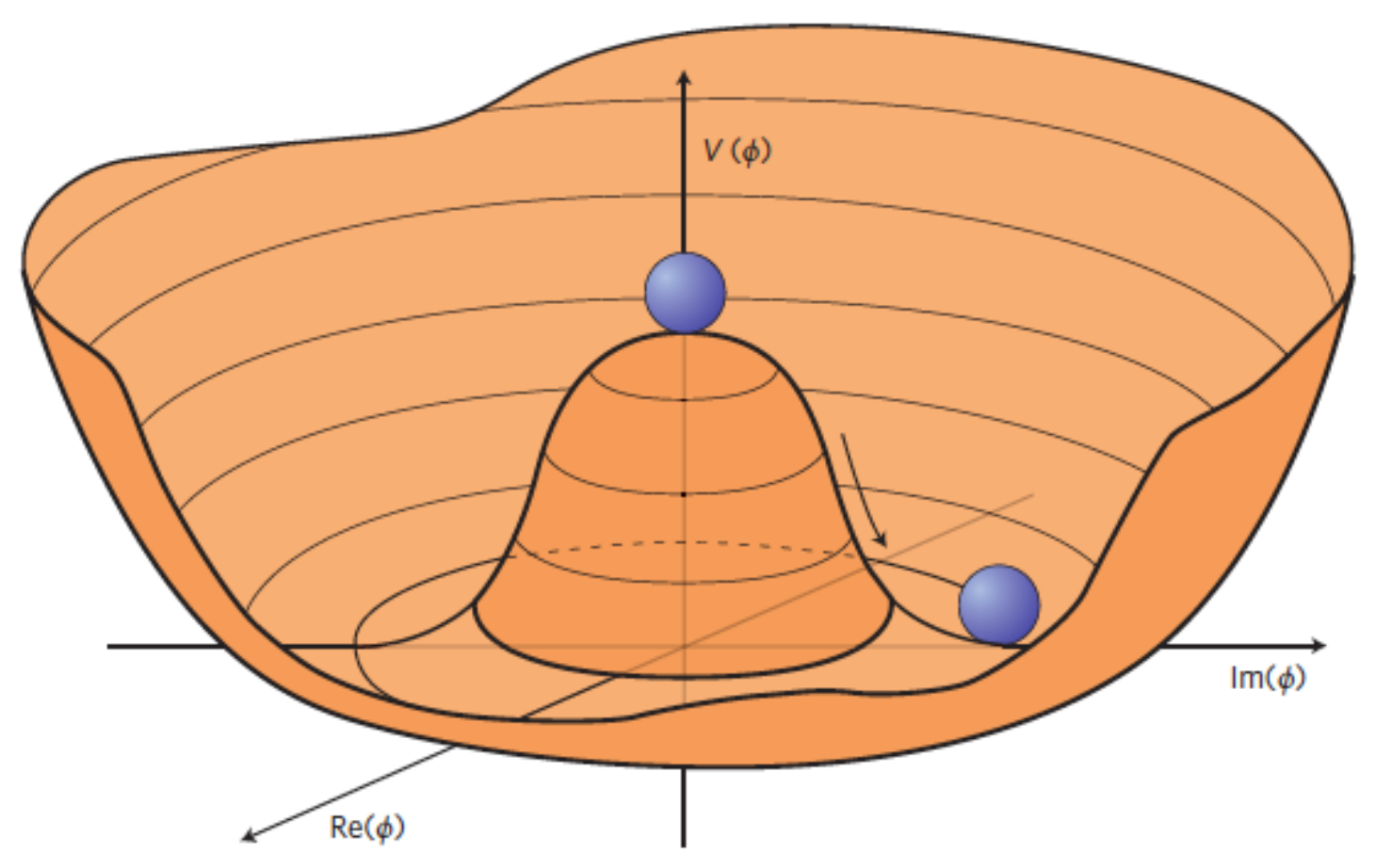}
\includegraphics[width=0.45\textwidth]{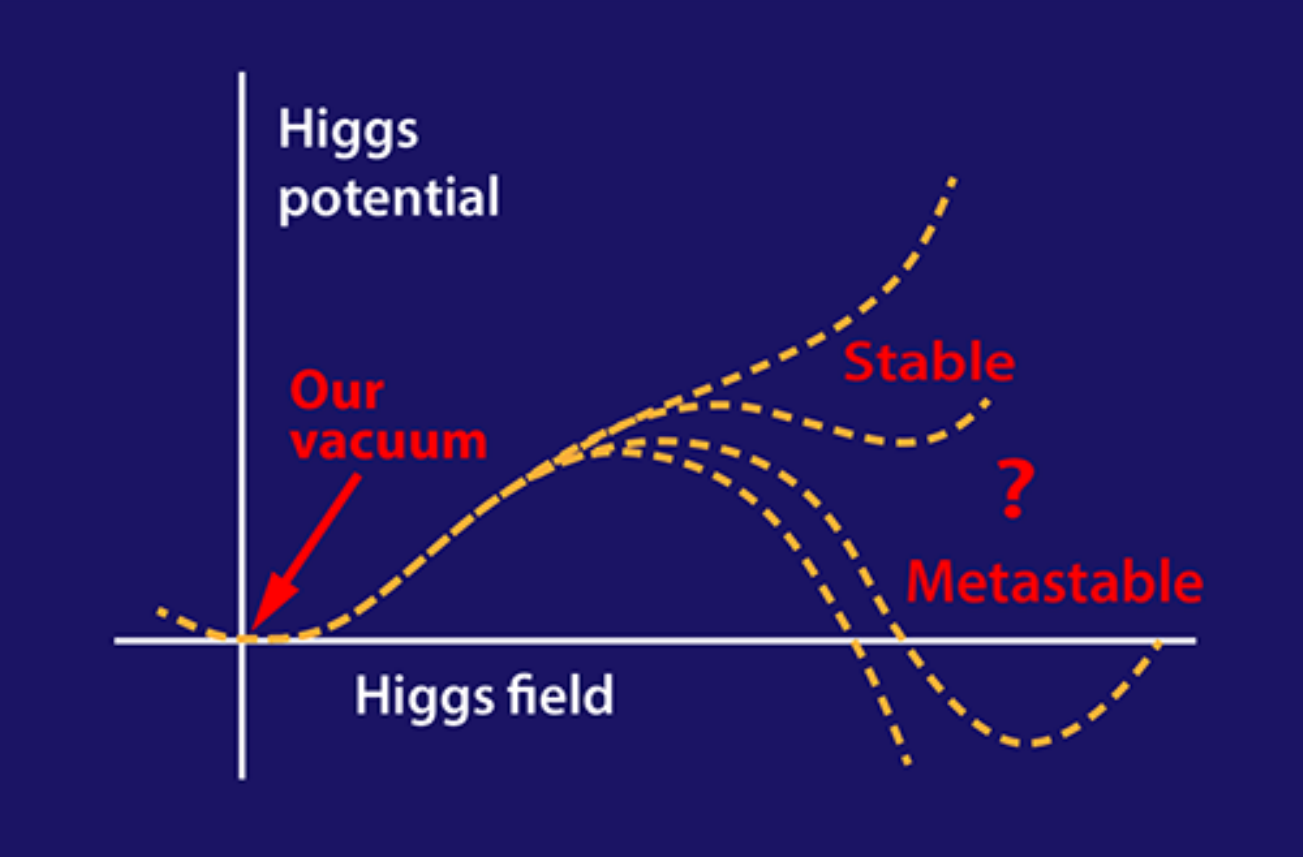}}
\caption{ 
Left: The Higgs potential for $\mu^2 <0$, 
Eq.~(\ref{eq:2b}).
Choosing any of the points at the bottom of the potential spontaneously breaks the rotational U(1) symmetry.
Right:
Quantum corrections can change the shape of the Higgs potential 
as discussed in Section \ref{sec:vac}. Here the minimum of ``our vacuum'' is taken at 
$|\phi| = \frac{v}{\sqrt{2}}$.
Figure from Ref.\cite{Kusenko:2015ab}. 
}%
\label{fig:higgs}
\end{figure}

To understand the BEH mechanism,
consider the coupling of the
gauge field ${\rm A_{\mu}}$ to a complex scalar field ${\rm \phi}$
via the gauge covariant derivative with coupling constant $g$, 
viz.
$
D_{\mu} {\rm \phi} =
\bigl[ \partial_{\mu} 
+ i g 
{\rm A_{\mu}} \bigr] \phi %
$.
The scalar field is taken 
with potential
\begin{equation}
V(\phi) = \frac{1}{2} \mu^2 \phi^2 + \frac{1}{4} \lambda \phi^4 .
\label{eq:2b}
\end{equation}
Here the self-coupling $\lambda \geq 0$ so the potential has a finite minimum, as required for vacuum stability.
If $\mu^2>0$ 
the potential 
describes
a particle with mass
$\mu$.
When 
$\mu^2 < 0$
the potential has a minimum at
\begin{equation}
| \phi | 
\equiv \frac{v}{\sqrt{2}}
=
\sqrt{- \frac{\mu^2}{2 \lambda}} .
\label{eq:2c}
\end{equation}
This is illustrated in Fig.~\ref{fig:higgs}.
Excitations around the 
degenerate minima of the potential -- the bottom of the ``Mexican hat'' --
correspond to a massless Goldstone state.
Gauge freedom allows us to choose
$v$ as the vacuum expectation value
of the real part of $\phi$
with all choices of vacuum states being degenerate
as well as 
physically equivalent.
Expanding the scalar field about this minimum, the Goldstone mode is 
``eaten''
to become the longitudinal 
mode of ${\rm A_{\mu}}$
which now acquires mass
$g^2 v^2$.
The Higgs boson H with mass squared
$m_{\rm H}^2 = 2 \lambda v^2$ corresponds to excitations up the rim of the potential.

The issue of massive gauge bosons
was first solved
by Anderson \cite{Anderson:1963pc}
in discussion of
massive ``photons'', called plasmons, 
in superconductors
\cite{Anderson:1963pc}
The photon behaves as a 
wave on a sea of BCS Cooper pairs which here act as the scalar field $\phi$, condensing in the ground state.
The order parameter is not rigid with zero momentum Cooper pairs
but fluctuates 
in the longitudinal 
component
to preserve the translational symmetry of the electron gas. 
The plasmon's transverse component is a modification of a real photon propagating in the plasma whereas the longitudinal mode is an attribute  of the system.
Massive plasmons are manifest through exponential decrease of the magnetic field inside the superconductor (the Meissner effect).

The relativistic case 
\cite{Higgs:1964ia,Higgs:1964pj,Higgs:1966ev,Englert:1964et}
has been introduced to provide a consistent model of
weak interactions
in particle physics~\cite{Glashow:1961tr,Weinberg:1967tq,Salam:1968rm,'tHooft:1972fi}. Contrary to the BCS case,  the weak interaction requires the introduction of an additional fundamental scalar field. A dynamic explanation of the Higgs mechanism {\it \`a la} BCS would be a major breakthrough and is one of the fundamental motivations to measure with the highest possible precision the properties of the Higgs particle.
For a more detailed history of theoretical developments, see
Ref.~\cite{Lykken:2013oca}.
For weak interactions the 
gauge group is SU(2). There are three massless Goldstone modes which combine to form the massive W
charged 
bosons and
the massive Z. 
The massless photon 
and neutral Z boson
are linear combinations of the 
neutral weak SU(2)  gauge boson and a U(1)
gauge boson called hypercharge.
Within the Standard Model,
the BEH mechanism is also important for 
fermion masses, something required by parity violation of weak interactions
\cite{Veltman:1997nm}.
The weak interaction gauge
bosons couple to SU(2) doublets of 
left-handed leptons and quarks, 
whereas right-handed fermions are weak interaction neutral.
Singlet mass terms 
for the charged fermions
are constructed by contracting the left-handed fermion doublets with the 
SU(2) Higgs doublet, including the vev, and then multiplying by the right handed fermion.
The Standard Model particle masses are
\begin{equation}
m_{\rm W}^2 = \frac{1}{4} g^2 v^2 , 
\ \ \ \ \
m_{\rm Z}^2 = \frac{1}{4} (g^2 + g'^2 )v^2 , 
\ \ \ \ \
m_{\rm f} = y_{\rm f} \frac{v}{\sqrt{2}} , 
\ \ \ \ \
m_{\rm H}^2 = 2 \lambda v^2
.
\label{eq:2d}
\end{equation}
Here $g$ and $g'$ are the SU(2) and U(1)
gauge couplings
and $y_{\rm f}$ denotes the fermion Yukawa coupling to the Higgs boson.
Before tiny neutrino masses
the Standard Model has 18 parameters: 
3 gauge couplings
and
15 in the Higgs sector
(6 quark masses, 3 charged leptons, 4 quark mixing angles including one CP violating complex phase, the W and Higgs masses).
There is a wide range of masses
with 
$m_{\rm W} = 80$ GeV, $m_{\rm Z} = 91$ GeV, $m_{\rm H} = 125$ GeV and the charged
fermion masses ranging from 0.5 MeV for the electron up to 173 GeV for the top quark.

Small changes in Higgs couplings and particle masses 
can lead to 
a very different Universe, assuming that the vacuum remains
stable, 
with one example that 
small changes in the light-quark masses can prevent
Big Bang nucleosynthesis \cite{Carr:1979sg}.
Once radiative corrections are taken into account - see Section \ref{sec:vac} - the stability of the Higgs vacuum is very sensitive to the value of the top quark mass.
Also and vitally, the Higgs boson cannot be too heavy to do its job with maintaining perturbative unitarity.
Indeed, if the Higgs boson had not been found at the LHC
new strong dynamics
would have been needed in the energy range of the experiments,
e.g. involving strongly interacting ${\rm W^+ W^-}$ scattering with the Higgs boson replaced by some broad resonance 
in the WW system \cite{Chanowitz:2004gk}.

In contrast to particle physics where the Higgs is treated as an elementary particle, 
in quantum condensed matter systems 
it forms as a collective mode 
\cite{PhysRevB.26.4883}.
Following the Higgs boson discovery in High Energy Physics,
collective Higgs states have recently been observed
in superconductors
\cite{Sherman:2015};
for discussion see
Refs.~\cite{Anderson:2015,Shimano:2020}.
In the particle physics Standard Model
one would like to understand deeper the more fundamental origin as well as any
internal structure of the Higgs boson.

\section{The Higgs boson discovery at CERN  and first 
measurements}
\label{Discovery}

The year 2012 was a seminal one for particle physics. More than 40 years after the 
original postulation for 
electroweak symmetry breaking via the 
BEH mechanism, the first potential experimental evidence for the  quantum excitation of that field was announced by the ATLAS and CMS experiments on July 4th, at a joint seminar held at CERN, the host of the LHC, and via video connection with the opening session at the international bi-yearly particle physics conference in Melbourne, Australia. 

The LHC is an atom smasher that collides proton beams, and has operated at centre of mass energy of 7 and 8 TeV 
(run~1; 2010-2012) and 13 TeV (run~2; 2015-2018) to search for new particles and phenomena ~\cite{Bruning:2004ej}.
ATLAS~\cite{Aad:2008zzm}
and CMS~\cite{Chatrchyan:2008aa} are two general purpose 
experiments making use of the highest luminosities at the LHC, and are its Higgs hunters.

The expected properties of a Standard Model Higgs boson are theoretically well known for any given -- but in 2012 unknown -- mass value of this particle. 
The announcement from ATLAS and CMS was based on the data collected and analysed until
then and enough for both experimental collaborations to claim independently the observation of a new particle, i.e.
the significance of the result was larger than five standard deviations, or 5$\sigma$, away from a 
background-only result, meaning that the chance of this
result being due to a fluctuation of the background is 
less than 1 in 3,500,000.
Measurements which give a significance above 3$\sigma$  are considered as evidence. A useful quantity extracted from the data is 
the "signal strength" which is the ratio of the 
signal rate divided by the 
 predicted rate for a Standard Model 
 Higgs boson at a given
 mass, and is denoted by the symbol $\mu_S$. The closer 
 $\mu_S$ is to one, the more it resembles a 
 Standard Model Higgs boson.

According to the Standard Model, 
a produced Higgs boson  has a lifetime of
 only $\sim 1.6 \times 10^{-22}$ seconds if
 its mass is about 125 GeV, after which it disintegrates into
 particles that are 
recorded by the experiments. Hence experimenters 
search for these footprints of the Higgs boson in the detectors.
The data showed that the new particle had a mass of around 125 GeV,  about 133 times the mass 
of a proton, and decayed into vector bosons, namely a pair of photons, W bosons, or Z bosons, exactly as anticipated from theory, and therefore got
labelled "a Higgs boson candidate". The observed decay into two photons  meant that this new particle could not have spin-one, due to the Landau-Yang theorem~\cite{Landau:1948kw,Yang:1950rg}.
The Higgs boson has no electric charge of its own and 
decays into two photons via a fermion or W boson loop.

The world-wide particle physics community was excited by the announcement, and 
the international press coverage of the event was huge.

A few months later a next crucial step was made by verifying the quantum properties of this new particle, demonstrating that it had to be a scalar spin-zero particle, as required for 
the messenger of the BEH field. 
While some small level mixing with a CP-odd component is 
still possible, the new particle has been firmly excluded to be a pure CP-odd state\cite{Chatrchyan:2012jja,Khachatryan:2014kca,Aad:2013xqa}.
This result promoted the particle to be called "a Higgs boson". 
This still allows that this new particle may not be the Standard Model Higgs boson, but a look-alike, such as a scalar  particle from an extended theory sector or even a composite particle.
For further insight the Higgs boson properties need to be mapped out in detail which presently can only be done at the LHC. Does this new particle also couple to the other known 
fundamental particles as expected, i.e. the  quarks and charged leptons? What is the exact mass value and width of this resonance?
Can we directly measure the shape of the Higgs field potential discussed in Section \ref{sec:higssth},  e.g. via Higgs boson pair production?

ATLAS and CMS discovered this new particle with a data sample of about 10 fb$^{-1}$ each
\cite{Foot2}.
An additional 15 fb$^{-1}$ was collected by the end of 2012 by both experiments in run~1. From 2015 to 2018, the LHC run~2, the
experiments collected 139 fb$^{-1}$ each at a higher center of mass energy: 13 TeV. In the next few
years the LHC will deliver for
each experiment  $\approx$ 150 fb$^{-1}$ in run~3 (2022-2024), and then the accelerator and both experiments will be upgraded
for the high luminosity phase to collect a total of
3-4 ab$^{-1}$ each.

Many analyses are presently still being finalized on the collected data set but 
several full run~2 results have been completed and show an emerging picture that we discuss next. In the collisions for this data set about 7 million Higgs bosons have been produced so the LHC
can be considered as the first Higgs factory, even though only a fraction can be identified and used to study the properties of the particle. 
Analyses of data from the LHC go in parallel with advances in precision theoretical calculations for the Standard Model production and decay rates as well as modelling of the 
backgrounds~\cite{Dawson:2018dcd,Heinrich:2020ybq,deFlorian:2016spz,deBlas:2019rxi}.

 In the Standard Model all the couplings of the Higgs boson to fermions and vector bosons were known 
 as a function of the Higgs boson's mass
 before the discovery of the Higgs boson. 
 The only parameter that was not predicted by the theory is the mass of the Higgs boson itself.
 Upper bounds could be obtained from unitarity in  longitudinal vector boson scattering, which was essential in convincing that the LHC should be able to either observe the Higgs boson or signs of new underlying strong dynamics in the TeV range. This is often referred to as the {\it no-lose} theorem.
 A 95\% confidence level upper bound on the
 Higgs boson mass of 166 GeV was already derived
 from previous electoweak measurements, mainly
 from the former LEP collider at CERN~\cite{Alcaraz:2006mx}.

 To measure the mass of the Higgs boson, decays into a pair of photons or Z bosons, with each Z boson itself decaying into a pair of electrons or muons, are the
channels {\it par excellence}: charged leptons and photons can be measured with
excellent precision in the LHC detectors and thus the mass of the Higgs boson can be fully determined from the invariant mass of the final state particles. 
The resulting distributions show 
typical resonant structures where the width of the resonance is 
determined by the detector resolution. The extracted
central mass value for the Higgs boson
from combined ATLAS and CMS
run~1 
measurements is reported in
Ref.~\cite{Aad:2015zhl}
to be 125.09 $\pm$ 
0.21 $\pm$ 0.11 GeV.
Subsequently new values for the mass are reported by 
CMS of 
125.38 $\pm$ 0.14 GeV~\cite{Sirunyan:2020xwk} 
and by ATLAS of
124.97 $\pm$ 0.24 GeV~\cite{Aaboud:2018wps}, 
with these measurements
using both the diphoton and ZZ decay channels.
It is remarkable that
we know the mass of this particle already to almost one per mille precision.

Interestingly, 
the mass of the Higgs boson at 
125 GeV is an "ideal choice" of Nature for a detailed
experimental study of this new particle. Indeed
the product of the branching ratios of the Standard Model Higgs boson in all decay channels available below the top-antitop threshold has been observed in~\cite{dEnterria:2012eip} 
to be a Gaussian distribution of the Higgs boson mass with a maximum centered at $m_{\rm H} \approx 125$ GeV, i.e. exactly at the mass value where this new boson has been discovered. No other Standard
Model Higgs boson mass value has a better combined signal-strength for the whole set of decay channels. Reversely, this mass value still allows 
for many Beyond the Standard Model scenarios.

\section{The Higgs boson couplings to fermions}
\label{fermions}

The Higgs boson was discovered in channels where it decays to gauge bosons. The observed inclusive production rate of the boson confirmed that the predicted main production process should be through gluon fusion
${\rm gg \to H}$.
This indirectly implied that the Higgs boson should couple to top quarks, involved in the decay quantum loop. However, direct evidence of the coupling of the Higgs boson to fermions is of paramount importance to demonstrate that the 
minimal
version of the Standard Model initially proposed is correct and 
that the same scalar field is responsible for the masses of the vector bosons and the charged fermions.

Such a direct test is establishing the decay of the Higgs boson into charged fermions,  and was a key physics target
for the LHC run~2. Decays to all fermions are kinematically allowed except 
for the  decay into a top anti-top quark pair, but we can still explore the 
Higgs boson to top quark
Yukawa coupling as reported in Section \ref{sec_top}. 
An important check is the quantitative comparison of the 
couplings strengths to the different fermions, which for the quantum particle associated with the BEH field 
in the Standard Model are
expected to be proportional to the masses of the fermions.

The most easily accessible channels are decays to the bottom or b-quarks and to the 
tau leptons, members of the third and most 
massive fermion generation.
First evidence for decays to the third fermion generation, tau leptons and
b-quarks, was reported already with run~1 data.  The couplings to the second fermion generation, 
the muon and the charm and strange quarks, are more challenging. The LHC is unlikely to be able to test couplings to first generation with the present methods at our disposal, and these are targets for  
a future very intense Higgs factory.

In this Section we next focus on the
Higgs boson coupling to the top quark. 
This is special
with the top being heavier than 
the Higgs boson
and with
top quark Yukawa coupling to the Higgs boson $y_{\rm t} \sim 1$.
We then discuss
masurements of the Higgs boson couplings
to the tau lepton, the bottom quark and the lighter mass fermions including a recent experimental highlight: observation of the Higgs boson to muon coupling.

\subsection{The Higgs boson to top quark Yukawa coupling}
\label{sec_top}

The top quark is the heaviest known fundamental fermion in Nature; its measured mass~\cite{Zyla:2020zbs} of $ 172.76 \pm 0.30$
means that the 
top quark Yukawa coupling is very large, close to one. 
The precise measurement of this coupling plays an  essential role in
the 
energy scale dependence
of the Higgs boson self-coupling,
which is essential
to
understanding the stability of the particle physics Higgs vacuum:
Does the current vacuum expectation value of the BEH
field correspond
to the real minimum of the Higgs potential?
(For more details, see Section \ref{sec:vac}.) 
It also allows 
for a fundamental check of the quantum consistency of the theory by comparison with indirect measurement through the main gluon fusion production process discussed above, 
which necessarily proceeds through quantum loop corrections and is therefore potentially sensitive to contributions from other yet unobserved states.

\begin{figure}[t!]  
\centerline
{\includegraphics[width=0.56\textwidth]
{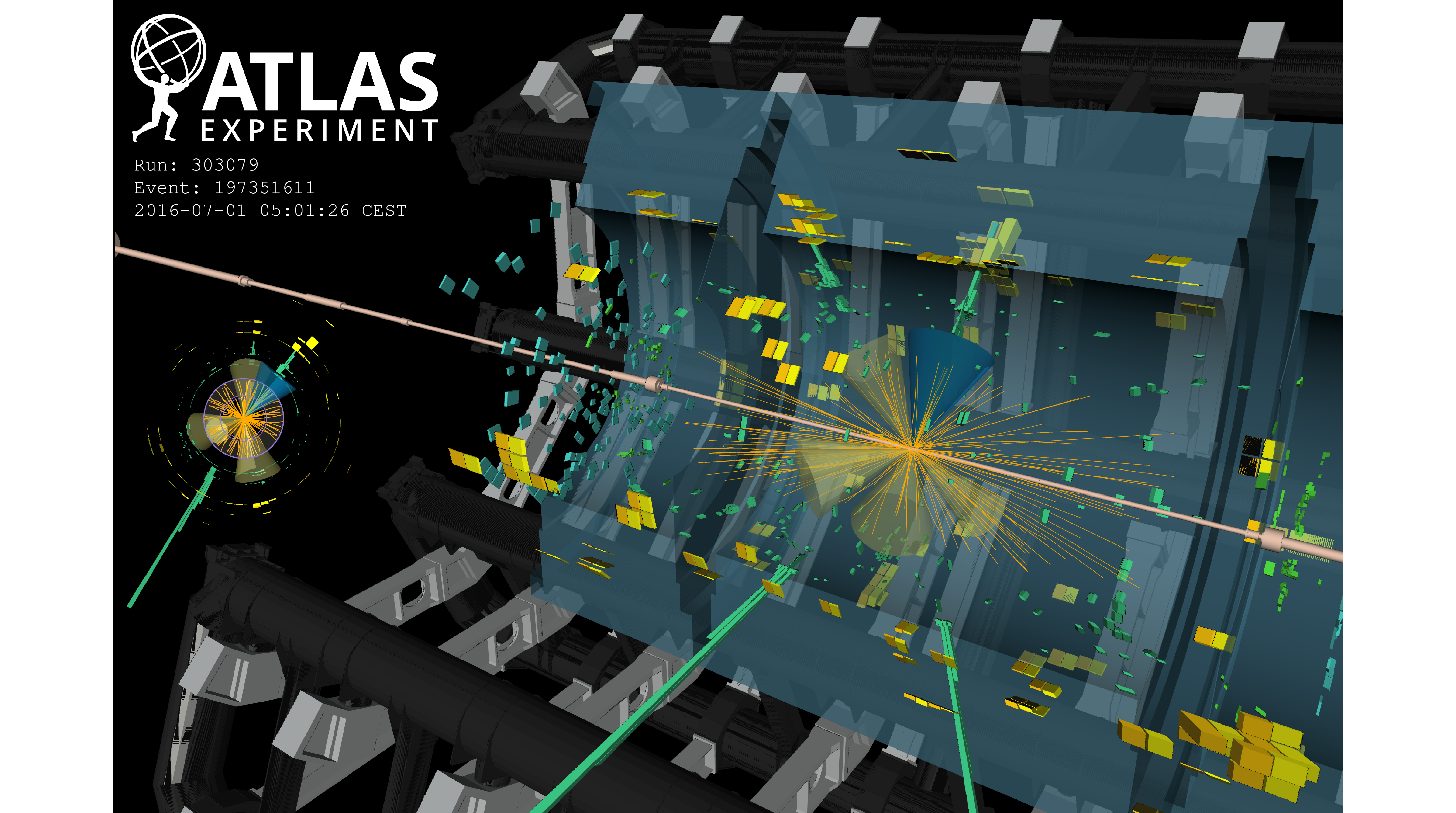}
\includegraphics[width=0.44\textwidth]
{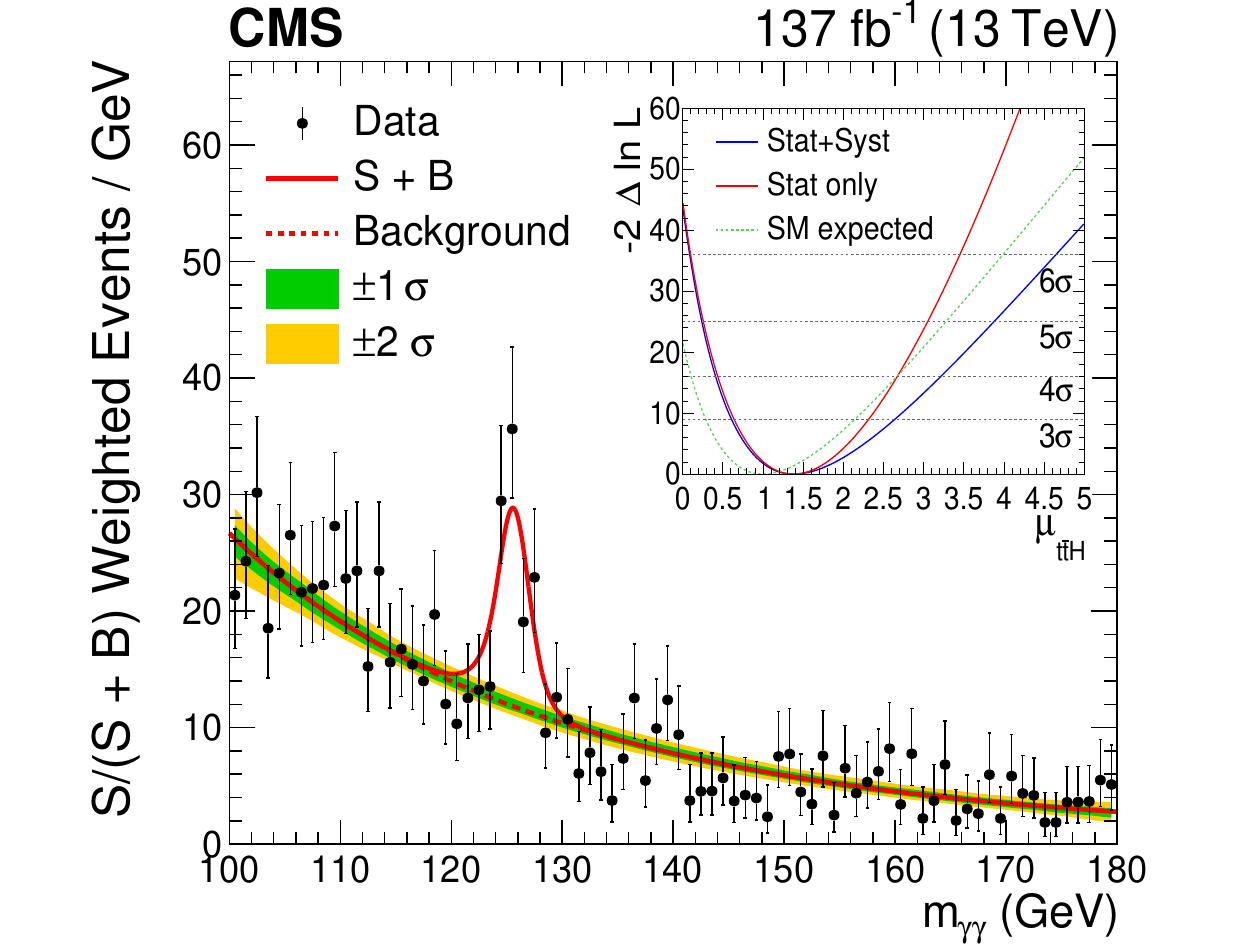}}
\caption{Left: A 3D event display of a candidate H$\rightarrow \gamma\gamma$ in the ${\rm pp \rightarrow t\overline{t}H}$ production mode, exemplifying the complex topologies of events where in addition of the two isolated photons (in green in the lower part of the detector), 6 jets are present among which one is tagged as originating from a $b$-quark (blue cone).
Right: The distribution of the invariant mass of the diphoton system for events selected in 
${\rm t\overline{t}}$H specific topologies.
The figure also displays the measurement's likelihood as a function of the strength of signal, indicating the subdominant impact of systematic uncertainties in this channel. Figure taken from Ref.~\cite{Sirunyan:2020sum}}
\label{fig:ttH-CP}
\end{figure}

A direct measurement can be made through the associated production of the Higgs boson with a pair of ${\rm t\overline{t}}$ quarks.
The topologies of events are complex and typically contain many jets, among which two at least originate from b-quarks, electrons or muons, and the decay products of the Higgs boson itself, 
see Fig.~\ref{fig:ttH-CP}
(left).
The first direct observation of the top Yukawa coupling of the Higgs boson was achieved only with a large but partial run 2 dataset and using all Higgs boson decay channels  ${\rm b\overline{b}, \tau^+\tau^-, WW^*, ZZ^*}$ and $\gamma\gamma$ by ATLAS and CMS~\cite{Aaboud:2018urx,Sirunyan:2018hoz}. The respective 
signal strengths for ttH production were found to be 
$1.32 \pm 0.39 $ (ATLAS)
and
$1.26\pm 0.31$ (CMS), in accord with the 
Standard Model expectation.

With the entire dataset, the diphoton channel alone provides an unambiguous observation of the $\rm pp \rightarrow t\overline{t}H$ production process~\cite{Sirunyan:2020sum, Aad:2020ivc}, 
see Fig.~\ref{fig:ttH-CP}
(right). 
The other Higgs boson decay channels are more challenging for precision measurements and
a key ingredient to improve their sensitivity 
relies on the progress in the theoretical predictions for the 
backgrounds.

The presence of the Higgs boson with a large top quark Yukawa coupling can also be indirectly measured through production processes where the Higgs boson does not appear in the final state but contributes as 
an exchange particle in the intermediate state of the reaction.
These indirect measurements have been carried out in top pair production processes, including the spectacular four-top channel for which first evidence has been observed, but are so far not competitive with the constraints from the direct observation of the associated production of a Higgs boson with a top quark pair.

Single top quark production together with a Higgs boson
has also been searched for in the experiments.
This process is especially interesting 
\cite{Barger:2018tqn}
since it is sensitive to the sign of the top quark Yukawa coupling $y_{\rm t}$ through the tree level interference between the production through the Higgs boson emission by a top quark and a W boson~\cite{Farina:2012xp}.
For Standard Model like couplings of the Higgs boson to W and Z bosons, CMS data~\cite{Sirunyan:2018lzm}
with 36 fb$^{-1}$
favours positive values of $y_{\rm t}$ and
exclude negative values below -0.9 $y_{\rm t}^{\rm SM}$.
Combined measurements
with integrated luminosity of 137 fb$^{-1}$
of 
t$\bar{\rm t}$H and tH
production 
in final states with electrons, muons and hadronically decaying tau leptons
yield constraints on
$\kappa_{\rm t} = y_{\rm t}/y_{\rm t}^{\rm SM}$ in the range
$-0.9 < \kappa_{\rm t} < -0.7$
or $0.7 < \kappa_{\rm t} < 1.1$~\cite{Sirunyan:2020icl}.

\subsection{The Higgs boson coupling to tau leptons }
Tau leptons have a mass of 1.777 GeV and for a Standard Model Higgs boson of 125 GeV the decay rate, or branching ratio, 
into a $\tau^- \tau^+$
lepton pair is about 6.3\%. Tau leptons are unstable, though, and decay with a 
mean lifetime of about 10$^{-13}$ seconds 
at the LHC 
into a narrow low-multiplicity hadronic jet, or a 
muon or electron, and in all cases in association with one or more neutrinos which go 
undetected in the experiments. The experiments have developed refined $\tau$-tagging methods in data for the 
most important $\tau$ decay channels, and have also been using additional event activity characteristics apart 
from the Higgs boson production to master and control the large backgrounds from non-Higgs boson  production processes. 
During run~1 
the Higgs boson to $\tau^- \tau^+$
decay was established by both experiments with a significance of about
3$\sigma$ for CMS and 4.5$\sigma$ for ATLAS~\cite{Chatrchyan:2014nva, Aad:2015vsa}
with the 5$\sigma$
threshold already crossed
in run~1 by
the ATLAS and CMS combination~\cite{Khachatryan:2016vau}.

Recent results based on partial run~2 data confirm these results and have established
an observation of the Higgs boson to $\tau^- \tau^+$
decay channel. 
CMS combined results of several production channels measured with 36 fb$^{-1}$ of data and
produced the overall result of an observation of the Higgs boson decaying into a $\tau^- \tau^+$ pair
with a significance of 5.5$\sigma$ and  signal strength for a Standard Model Higgs boson $\mu_S$ of 
 $ 1.24\pm^{0.29}_{0.27} $
 from run~2 data~\cite{Sirunyan:2018cpi} 
 or
 $0.98 \pm 0.18$
 with a combined
 significance of 
 5.9$\sigma$
 when 
 both run~1 and run~2 results are
 included
 \cite{Sirunyan:2017khh}.
The ATLAS result based on 36 fb$^{-1}$ of run~2 data
and
run~1 results
shows a significance of 6.4$\sigma$~\cite{Aaboud:2018pen} and
measures a cross section in accord with the 125 GeV Higgs boson prediction.
Overall the decay rate for 
$\rm H\rightarrow \tau^- \tau^+$ is found to be very close to the 
one  expected for a  Higgs boson with a mass of 125 GeV.

\begin{figure}[t!]  
\centerline
{\includegraphics[width=0.58\textwidth]
{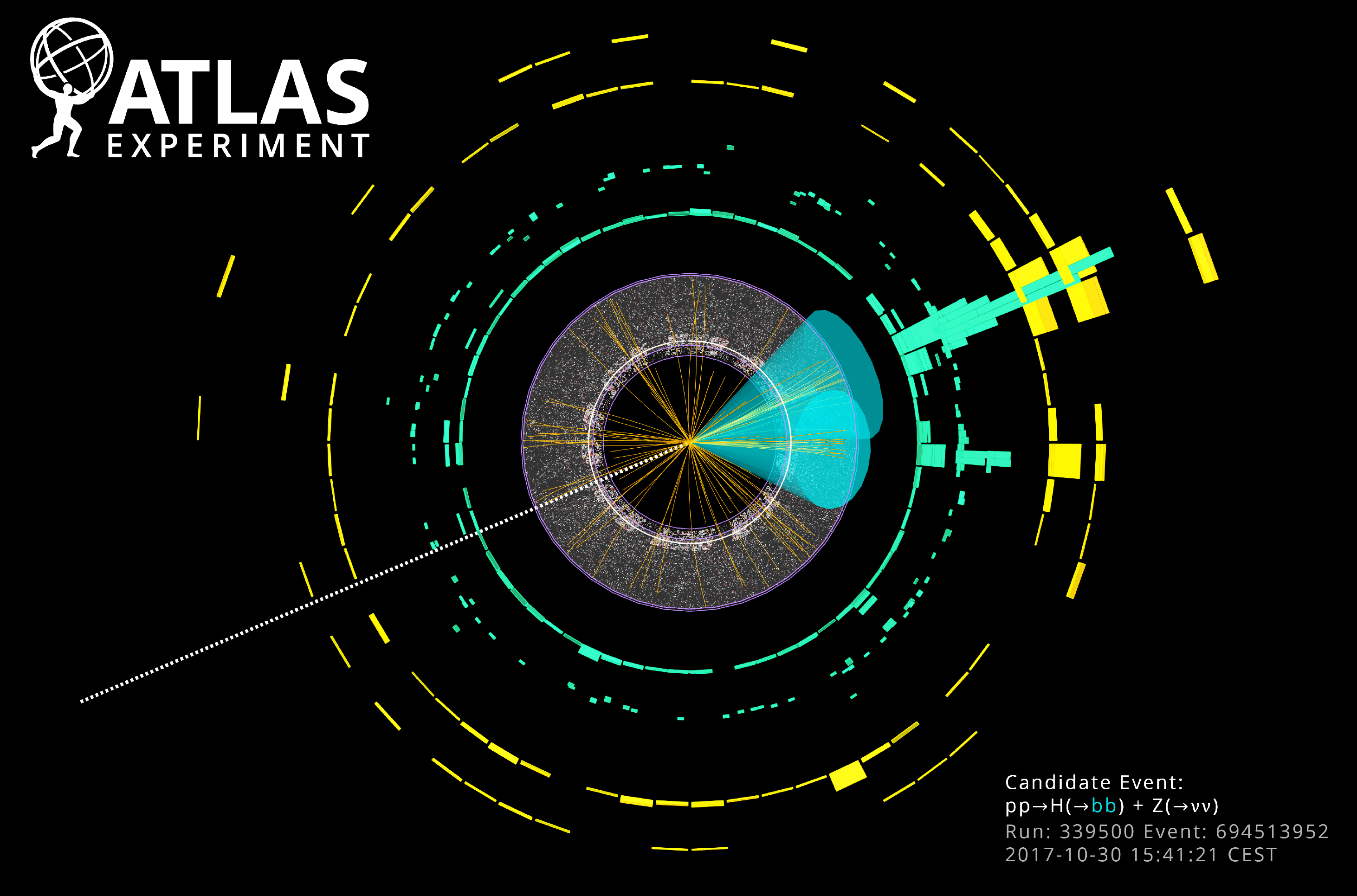}
\includegraphics[width=0.40\textwidth]
{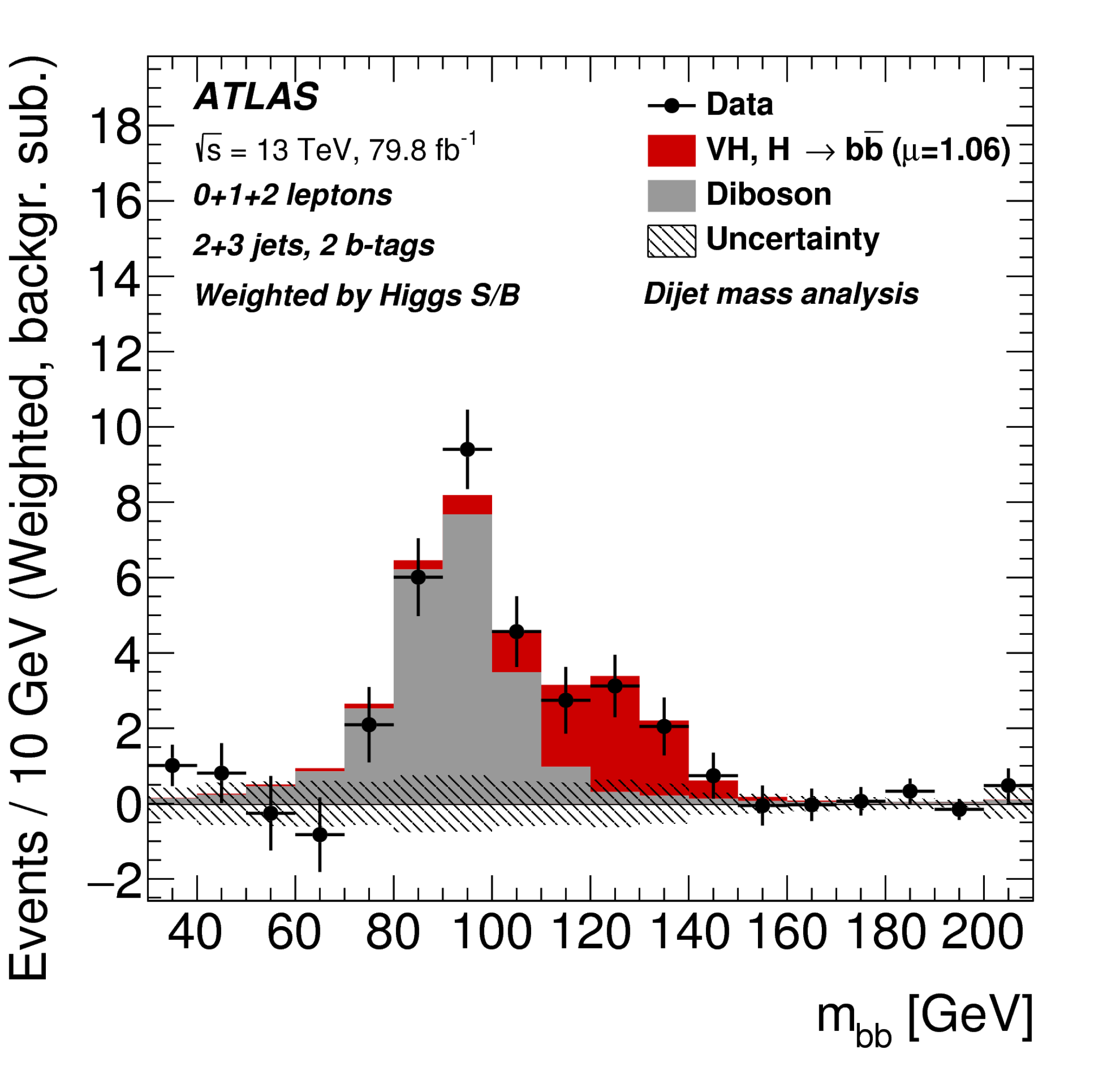}}
\caption{ An event display of a candidate $\rm{H}\rightarrow \rm{b}
\overline{\rm{b}}$ decay in the plane transverse to the beam axis (left); the cones illustrate the two reconstructed jets of the b-quark decays. The distribution of the $m_{\rm bb}$ invariant mass, showing the $\rm Z \rightarrow \rm{b} 
\overline{\rm{b}} $ (grey) and $\rm{H}\rightarrow \rm{b}
\overline{\rm{b}}$ signal (red) signal (right). Figure taken from Ref.~\cite{Aaboud:2018zhk}}%
\label{fig:bb}
\end{figure}

\subsection{The Higgs boson coupling to the bottom quarks}
The bottom  quark is the heaviest quark accessible in Higgs boson decays, and has a scheme
dependent mass of 4.2 GeV~\cite{Zyla:2020zbs}. Free quarks are not observable in Nature. Instead, at the LHC, quarks hadronize in jets of particles resulting from the colour force  that connects the produced 
quarks and breaks-up into
colourless hadrons, dominantly mesons. Hadrons containing a b-(anti)quark have short lifetimes, 
typically around 10$^{-12}$ seconds, {and thus cross 
distances in the detector of typically a few millimeters to centimeters}, 
a feature that can be efficiently explored by the experiments to tag particle jets that contain a b-quark. The branching ratio for a 125 
GeV  Higgs boson into a b anti-b quark pair is 58\%  and constitutes the largest Higgs boson decay channel at this mass value.
However, the cross sections of b anti-b quarks produced by Standard Model background processes  
are seven orders of magnitude larger than the Higgs boson production cross section, and hence largely dominate the search regions. 

The experimenters have been able 
 to spectacularly reduce these backgrounds by selecting special kinematic 
regions and by using additional event information, such as
additional associated 
jets and in particular heavy vector W and Z bosons (where the W and Z bosons decay leptonically), to extract the Higgs boson to bottom quark decay signal. In 2018 both experiments announced the observation of this decay
channel. 
ATLAS
reported a $\rm{H}\rightarrow \rm{b}
\overline{\rm{b}}$ decay signal with
5.4$\sigma$ and a 
signal strength of $1.01\pm 0.20$ ~\cite{Aaboud:2018zhk} based on up to 79.8 fb$^{-1}$ of run~2
and about 25 fb$^{-1}$ run~1 data.
CMS observed this channel with
5.6$\sigma$ significance and a 
signal strength of $1.04\pm 0.20$ ~\cite{Sirunyan:2018kst} based on 41.3fb$^{-1}$  run~2
and about 25 fb$^{-1}$ run~1 data.
Hence, the $\rm{H}\rightarrow \rm{b}
\overline{\rm{b}}$
 decay rate is consistent with the expectations for a 125 GeV Higgs boson.
Fig.~\ref{fig:bb} shows an event display of a candidate $\rm{H}\rightarrow \rm{b}
\overline{\rm{b}}$ decay (left), and (right) the distribution of the $m_{\rm bb}$ invariant mass, showing the $\rm Z \rightarrow \rm{b} 
\overline{\rm{b}} $ (grey) and $\rm{H}\rightarrow \rm{b}
\overline{\rm{b}}$ (red) signal.

\begin{figure}[t!]  
\centerline
{\includegraphics[width=0.54\textwidth]
{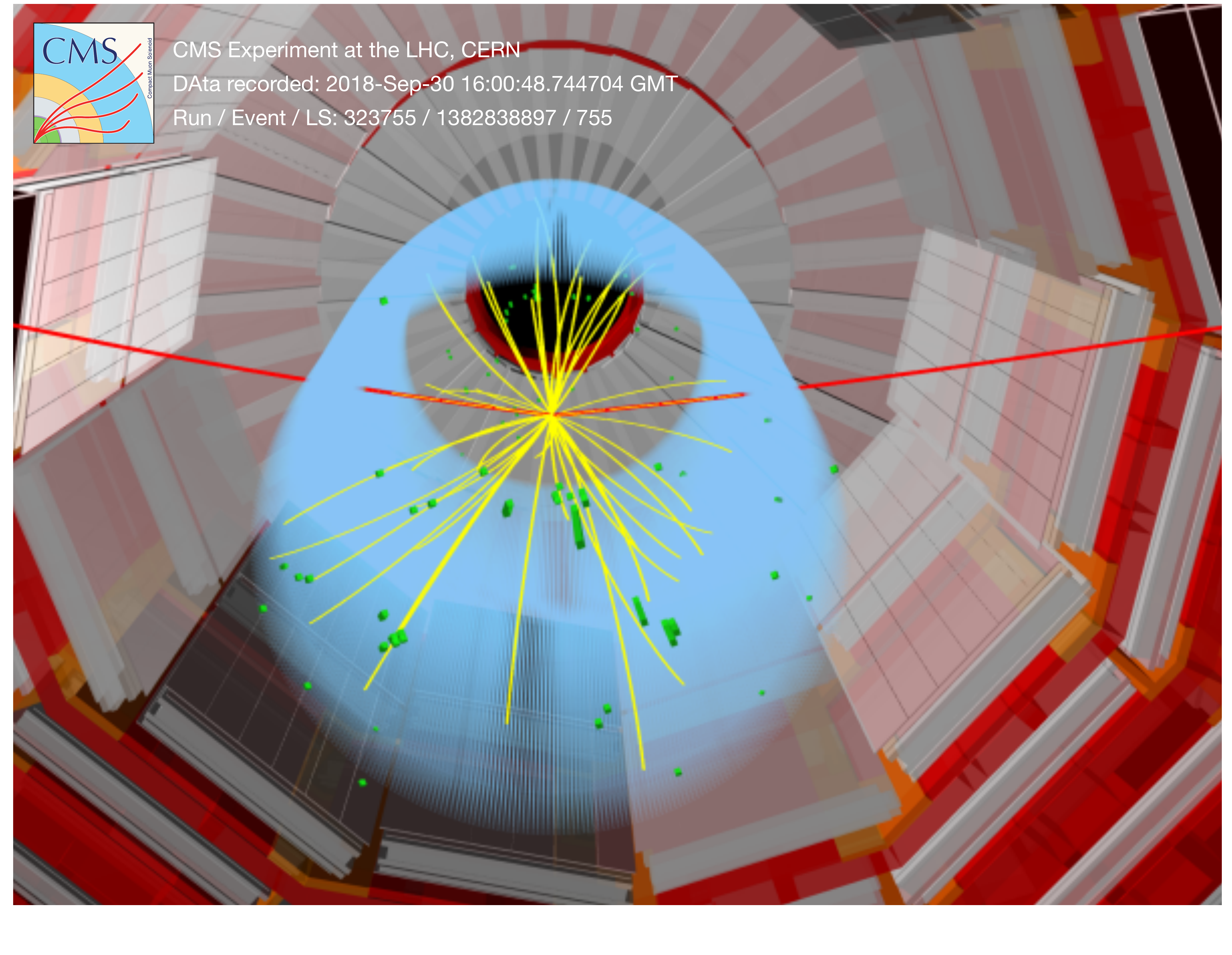}
\includegraphics[width=0.42\textwidth]
{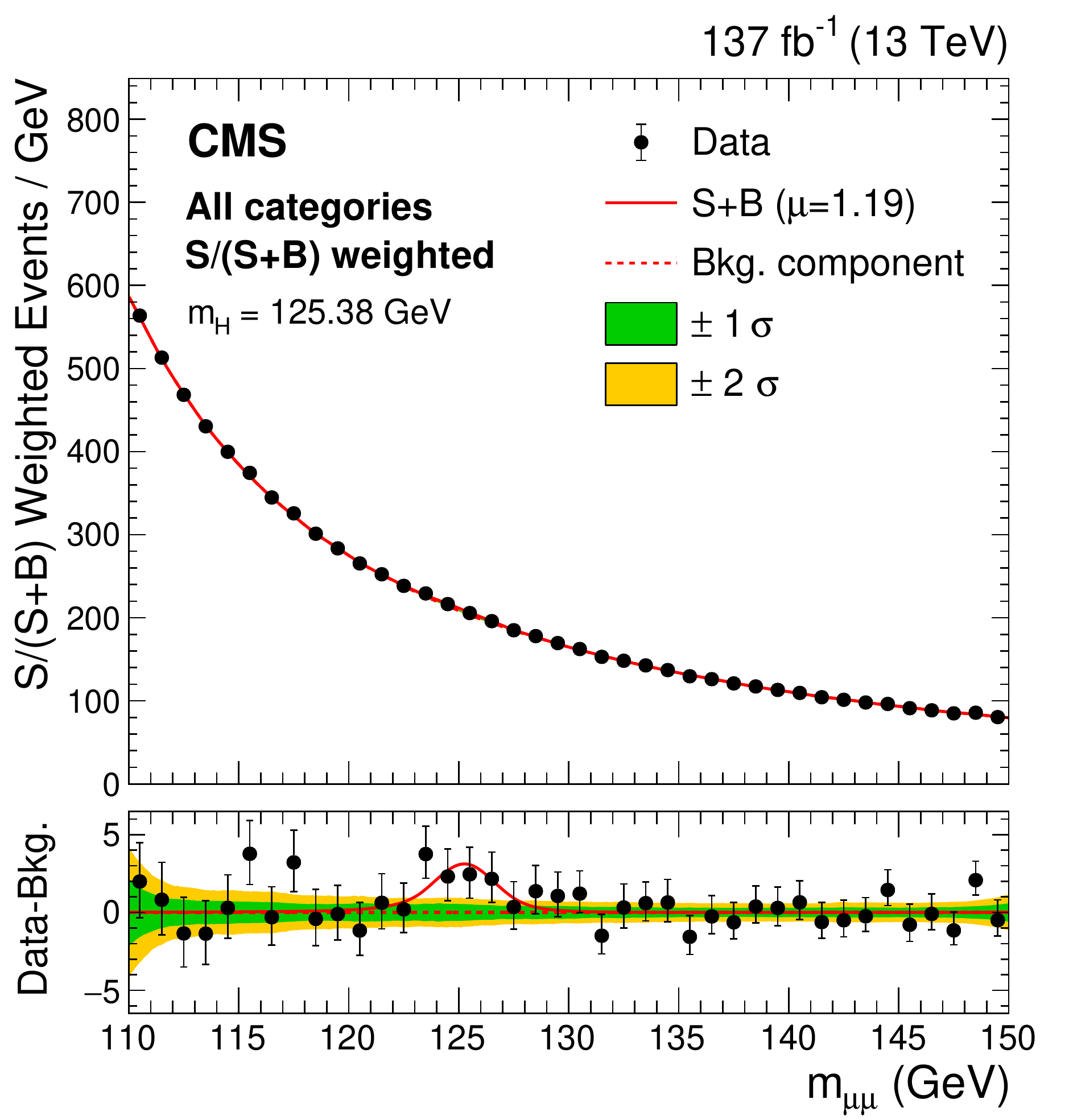}}
\caption{ A 3D event display of a candidate $\rm{H}\rightarrow \mu^- \mu^+$ decay illustrating the reconstructed trajectories of the two muons in red (left); the reconstructed invariant mass of the muon pair is consistent with that of the Higgs boson. The distribution of the $m_{\rm \mu \mu }$ invariant mass, and data $-$ background distribution (right). Figure taken from~\cite{Sirunyan:2020two}. }%
\label{fig:mumu}
\end{figure}

\subsection{The Higgs boson coupling to muons}
With the Higgs boson decay channels to the third generation fermions, b-quarks and tau leptons, 
now firmly established the natural next questions is: what about the second generation
fermions? Since these particles have lower masses and the signals are often subject to larger backgrounds, 
extracting 
these from the data gets increasingly challenging. 
However
these measurements are of uttermost importance to consolidate the scenario of the long-sought BEH mechanism.

The branching ratio of the $\rm H\rightarrow \mu^-\mu^+$ decay in the Standard Model for a Higgs boson of 125 GeV is
small, namely 2.18$ \times 10^{-4}$, but the final state is very simple: it amounts to a search for 
two oppositely charged muons which have large transverse  momenta, of the order of several 10's of
GeVs in the laboratory frame, and can be efficiently selected and reconstructed by the experiments. 
The signal resides on a large background tail: 
the Z boson
to muon pair cross section is 5 orders of magnitude larger than the expected signal.

The hunt for the $\rm{H}\rightarrow \mu^-\mu^+$ decays started early on. 
Any observed signal would have 
unexpected with the initial LHC luminosity since for a true Higgs particle this decay is expected to be strongly suppressed compared to, e.g., the Higgs boson to 
$\tau^- \tau^*$ decay,
and
would have been 
evidence that the newly found particle 
was {\it not} the Standard Model Higgs boson! 
And indeed,
no evidence for this decay was found in the run~1 data.

The full set of the run~2 data was recently used by ATLAS and CMS to search for this channel. The large background of mostly Drell-Yan di-muon production required the experiments to use 
sophisticated tools such as machine learning algorithms to extract a significant signal.
In summer 2020, after considerable effort, the experiments were successful and could report for 
the first time evidence for  $\rm{H}\rightarrow \mu^-\mu^+$ production.
CMS reported an observed significance of 3$\sigma$ and a signal strength of 
$1.19\pm 0.43$~\cite{Sirunyan:2020two}.
ATLAS reported an observed significance of 2$\sigma$ and a signal strength of 
$1.2\pm 0.6$   \cite{Aad:2020xfq}.
Fig.~\ref{fig:mumu} 
shows  an event display of a candidate $\rm{H}\rightarrow \mu^-\mu^+$ decay (left), and (right) the distribution of the $m_{\mu \mu}$ invariant mass. 

Clearly, these results open a new research program for the 
Higgs boson at the LHC: the detailed study of the second fermion
generation couplings to the Higgs boson. Presently, the precision of the results is statistics limited,
a handicap that will be remedied with the advent of 
the high 
luminosity run of the LHC expected to start well
before the end of the decade.

\subsection{The Higgs boson couplings to lighter quarks}
The quarks of the second fermion generation  have a scheme dependent mass of 1.27 GeV  for the 
charm quark, c,
and a current quark mass of $\approx 90$ MeV for the strange
quark, s~\cite{Zyla:2020zbs}.
The channel $\rm H\rightarrow c\overline{c}$ has a branching ratio of 2.8\% for a 125 GeV 
Higgs boson, a much larger background than for the $\rm H\rightarrow b\overline{b}$ decay channel and a less efficient
charm tagger compared to bottom quarks, and as a consequence is far more challenging to observe.

Direct searches for $\rm H\rightarrow c\overline{c}$ have been performed by the experiments, but
presently no evidence for this process can be reported, see e.g.  ATLAS\cite{Aaboud:2018fhh}. The
most sensitive result to date is an experimental sensitivity of a 
factor 70 (CMS\cite{Sirunyan:2019qia}) above 
the predicted Standard Model values.
Additional data, improved charm tagging and reconstruction efficiencies, and a deeper use of machine learning techniques
will no doubt push these sensitivities closer to the Standard Model observable limits but at this point we cannot yet 
be sure that this channel will become detectable at the LHC. Other channels are pursued as well, such as $\rm H \rightarrow J/\Psi \gamma$ and $\rm H \rightarrow \Psi (2S) \gamma$, 
the associated production of the Higgs boson with a c-quark, and the measurement of the charge asymmetry in the associated production mode of a Higgs boson with a vector boson.

The hunt for decays of the Higgs boson to lighter  
quarks and leptons is even more challenging
and is not expected to yield 
detectable signals
for Standard Model Higgs boson couplings at the LHC, but these decays have 
nevertheless been
searched for in topologies such as the di-electron  or the VZ and V$\gamma$ decay channels, with V a vector meson.

In all, wherever the LHC presently has 
sufficient sensitivity, its data 
dramatically show that the couplings to fermions are in agreement
with the expectations of the predictions
from the Standard Model BEH mechanism.
While the Higgs boson's couplings to light fermions are 
presently experimentally unknown, it has already been established that these couplings cannot be the same for all generations.

\section{Summary of measured Higgs boson's properties}
\label{properties}

In a few years the LHC will conclude its first phase of operation before a major luminosity upgrade for the machine and experiments is scheduled, 
but it is already instructive to summarize  
the overall emerging picture so far on this newly found particle. 
The
experiments have combined measurements of Higgs boson production cross sections and branching fractions. These combinations are based on the analyses of the Higgs boson decay modes $\rm H \rightarrow 
\gamma\gamma, ZZ , WW , \tau\tau ,
b\overline{b} , \mu \mu $, searches for decays into invisible final states, and on measurements of off-shell Higgs boson production. 
Such combinations are made by the experimental collaborations typically after all individual channels have been analysed for a large
fraction of the recorded data.
ATLAS (CMS) has produced such a combination based 
using up to 79.8 fb$^{-1}$ (35.9 
fb$^{-1}$) of proton-proton collision data\cite{Aad:2019mbh,Sirunyan:2018koj} collected  in run~2 and found overall global signal strength of the combined fit of all channels to be $\mu_S =1.11\pm^{0.09}_{0.08}$ ($\mu_S =1.17 \pm {0.10}$), i.e. close to one, the expected value in the Standard Model.
The final combined run~2 analysis based on 139fb$^{-1}$ from each experiment
will probably be available towards the end of 2021.

\begin{figure}[b!]  
\centerline
{\includegraphics[width=0.50\textwidth]
{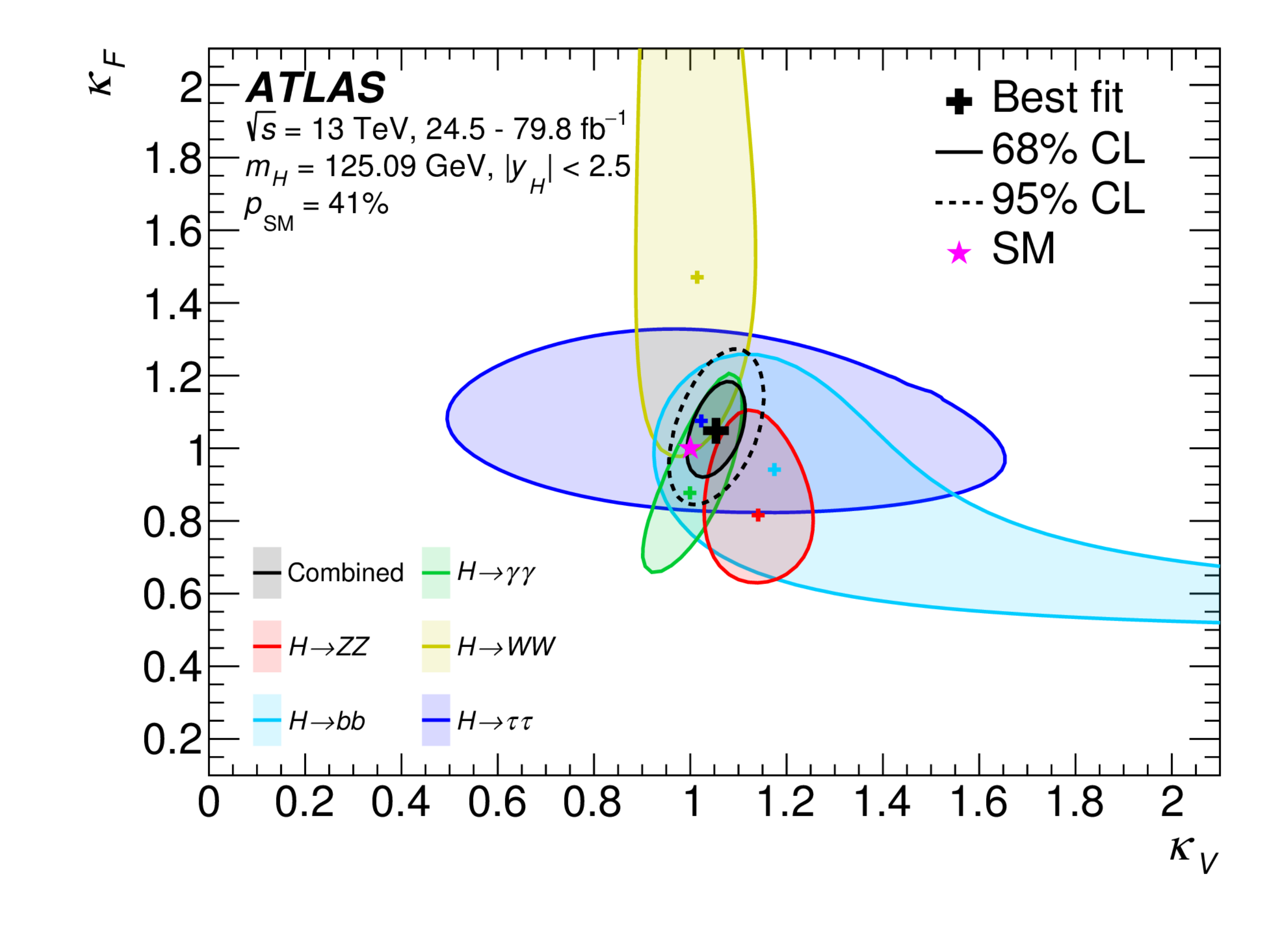}
\includegraphics[width=0.50\textwidth]
{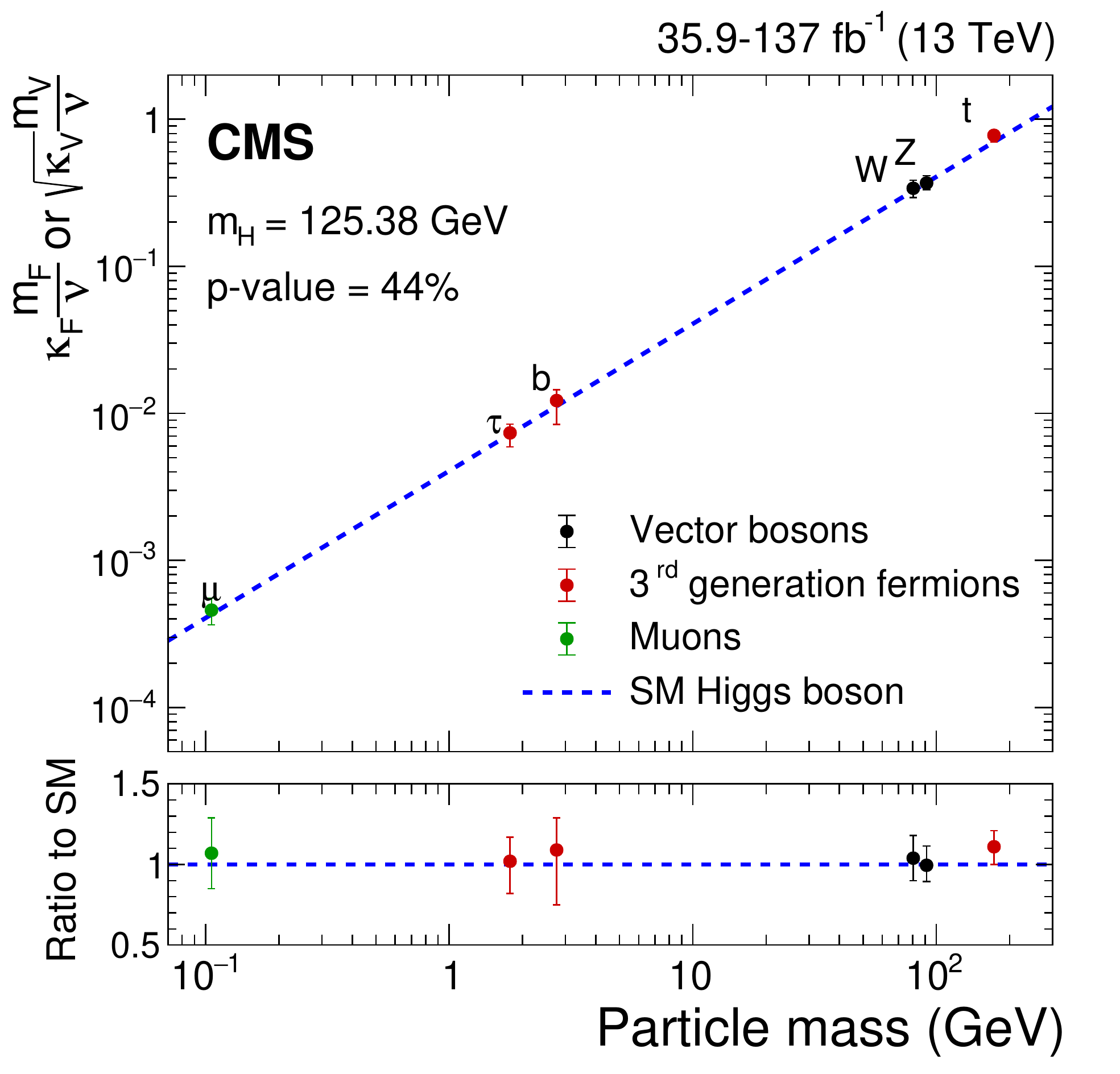}}
\caption{ Left: 
Negative log-likelihood contours at 68\% and 95\% CL in the ($\kappa_{\rm V} ,\kappa_{\rm F}$) plane for the individual decay modes and their combination assuming the coupling strengths to fermions and vector bosons to be positive. No contributions from invisible or undetected Higgs boson decays are assumed. The best-fit value for each measurement is indicated by a cross while the Standard Model hypothesis is indicated by a star. Right: the best fit estimates for the reduced coupling modifiers extracted for fermions and weak bosons from the resolved $\kappa$-framework compared to their corresponding prediction from the 
Standard Model. The error bars represent 68\% CL intervals for the measured parameters. In the lower panel, the ratios of the measured coupling modifiers values to their Standard Model predictions are shown. Figures taken from Ref.~\cite{Aad:2019mbh} and Ref.~\cite{Sirunyan:2020two}}.
\label{fig:coupling_mass}
\end{figure}

 These results are interpreted in terms of 
 so-called coupling modifiers $\kappa$ applied to the Standard Model couplings of the Higgs boson to other particles. 
 The coupling modifiers are derived from global fits to all the measurements in different production and
 decay channels assuming Standard Model relations between the channels, and therefore these $\kappa$ values
 do not measure the couplings directly but show levels of deviation from the Standard Model expectations.
 We can test the  coupling-strength scale factors.
 The result is shown for the different 
 decay channels in
 Fig.~\ref{fig:coupling_mass} (left) and shows
 the consistency of the couplings of the  Higgs boson to 
 vector bosons and fermions. 
 An overall fit 
 gives $\kappa_{\rm V} =1.05\pm0.04$ and 
 $\kappa_{\rm F} =1.05\pm0.09$, i.e. values close to one, the Standard Model prediction. The precision achieved so far in these measurements is not only relying on the excellent performance of the machine and the experiments, but also on the remarkable progress made in the theoretical predictions of the processes at stake, their simulation and their efficient reconstruction. 

 One of the most prominent achievements to date is 
 measurement of the hierarchy  of 
 the relative coupling strengths of the 
 different particles to the Higgs boson!
 In the Standard Model, the Yukawa coupling between the Higgs boson and the fermions, $y_{\rm F}$, is proportional
to the fermion masses $m_{\rm F}$, while the coupling to weak bosons 
is proportional to the square
of the vector boson masses $m_{\rm V}$,
with the latter following from the W and Z 
coupling to the Higgs boson via the Standard Model gauge covariant derivative~\cite{Altarelli:2013tya,Pokorski:1987ed,Aitchison:2004cs}.
 These relations are demonstrated by the data
 in a dramatic
 way in Fig.~\ref{fig:coupling_mass} 
 (right). 
 These are the fruits of 
 the first 10 years of data taking and 
 careful analysis at the LHC!
 
 The other landmark result of this first phase of the LHC is the fact that so far 
 no new phenomena
 beyond those predicted by the Standard Model have been observed.
This has led the experiments and the theory communities to perform combined interpretations of all measurements in Higgs, electroweak and top physics in a framework where the Standard Model is considered as an effective field theory~\cite{Giudice:2007fh,Grzadkowski:2010es,Ellis:2014dva,Falkowski:2014tna,Dawson:2020oco,deFlorian:2016spz}.

With the full LHC dataset, the precision on the couplings of the Higgs boson is expected to reach between 1\%\ and 2\%\ for the ones to gauge bosons and between 2\%\ and 4\%\ for couplings to the charged fermions of the third generation. Good precision will also be reached for the Higgs boson to muons 
 coupling and to  ${\rm Z\gamma}$ pairs~\cite{Cepeda:2019klc}. The final
 precision for Higgs boson coupling measurements at the LHC in the future will mostly be limited by the precision of theoretical predictions of signal and background processes.

The width
of the Higgs particle, which is a measure of its life-time and expected to be 4.1 MeV,  cannot be extracted from the observed experimental resonance line shape due to limited experimental resolution of the detectors.
Instead  an indirect method is used that compares the production rate of the 
on-mass shell Higgs boson to the production of the Higgs boson far off-mass shell, where it acts as a propagator in the production of a pair of vector bosons (W and Z, both on-mass shell).
While the on-mass shell rate depends on the Higgs boson's width, the 
off-mass shell rates does not and so comparing the rates of the two regimes gives an estimate of the Higgs boson natural width.
This method assumes that the running of the Higgs boson couplings do not deviate significantly from those expected from the Standard Model.
CMS extracts with 80.2 fb$^{-1}$ of data a central value of the width to be constrained to  3.2$\pm^{2.8}_{2.2}$ MeV at 68\% CL, and a range 
constrained between [0.08,9.16] MeV at 95\% CL\cite{Sirunyan:2019twz}.
The ATLAS experiment reports an upper limit of 14.4 MeV, based on 36 fb$^{-1}$ of 
data~\cite{Aaboud:2018puo}.
It is quite interesting that new results show now
also a lower limit of the
allowed range, but the measurement is presently still statistics limited. 
Since the ultimate LHC data 
sample will have an 20-fold higher statistics, this method is most promising
to experimentally verify the Higgs boson's width in the next 15 years.
Note however that there is a model dependent assumption using
this method that no extra new particles contribute to the
off-shell mass rate, i.e. to the rate of Higgs 
particles with mass larger than 180 GeV,
which would invalidate this width extraction.

\begin{figure}[t!]  
\centerline
{\includegraphics[width=0.65\textwidth]
{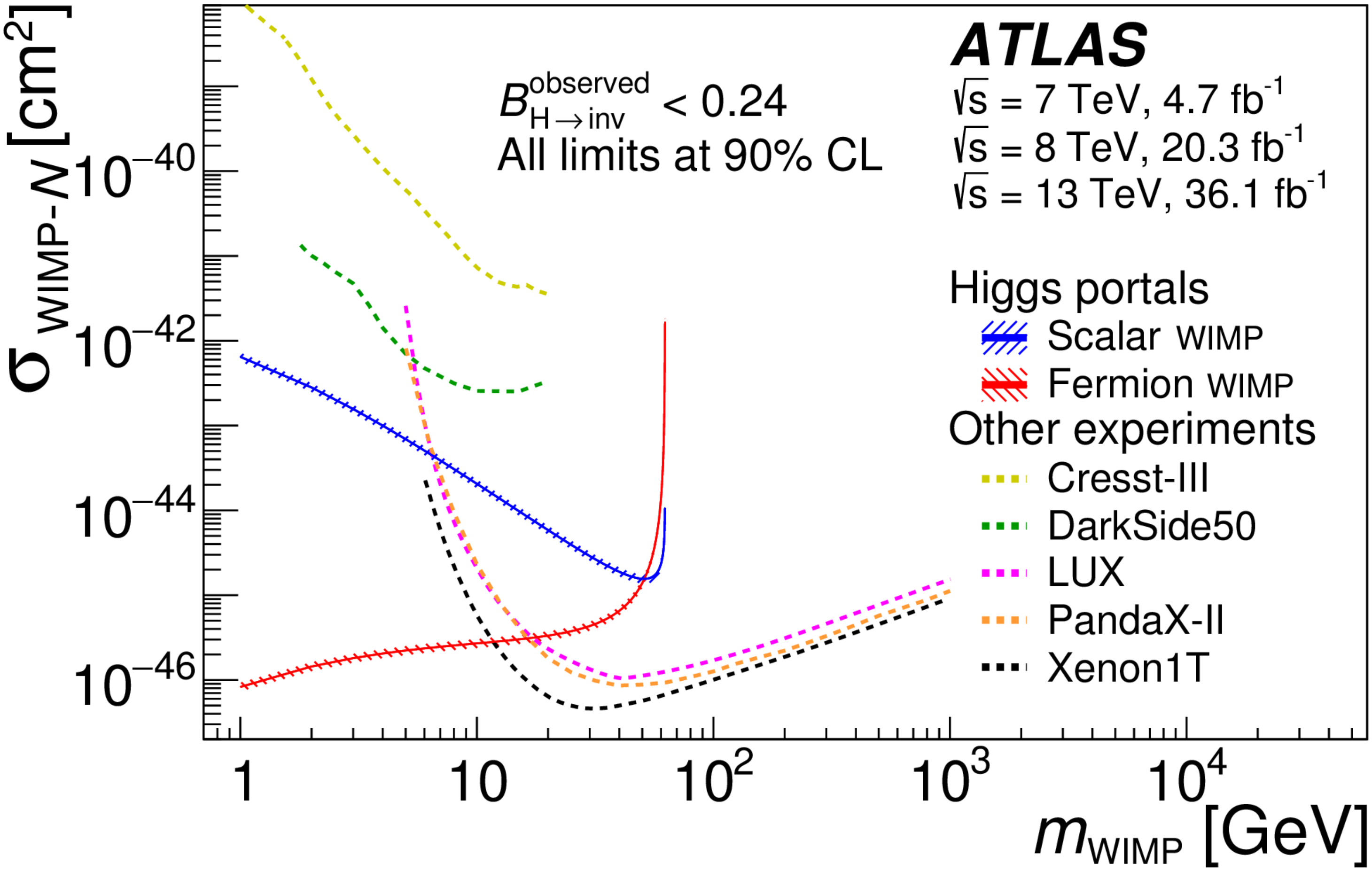}
}
\caption{Comparison of the upper limits at 90\% CL from direct detection experiments
\cite{Petricca:2017zdp, Akerib:2016vxi,Cui:2017nnn,
Aprile:2018dbl,Agnes:2018ves} on the spin-independent WIMP-nucleon scattering cross section to the observed exclusion limits, assuming Higgs portal scenarios where the 125 GeV Higgs boson decays to a pair of DM particles~\cite{Patt:2006fw}. The regions above the limit contours are excluded in the range shown in the plot. Figure taken from Ref.~\cite{Aaboud:2019rtt}}%
\label{fig:Higgs_DM}
\end{figure}

Pinning down the Higgs boson's width allows us to probe beyond the
Standard Model: it  accounts for possible
invisible decays 
into particles that do not interact in the 
detectors, e.g. such as the dark matter candidates
discussed in 
Refs.~\cite{Arcadi:2019lka,Carmona:2020uqx}.
In the Standard Model such decays are expected from neutrinos in the final state (from Z boson decays) and are rare, with a branching fraction of approximately $10^{-3}$. 
Invisible Higgs boson decays can be directly 
searched for 
in event topologies with significant missing transverse momentum. Both ATLAS and CMS have performed searches for these decays in all the main production modes 
yielding already stringent constraints on the invisible decay width~\cite{Sirunyan:2018owy} of $\sim$20\%. 
This can be turned into limits on dark matter
searches as shown \cite{Aaboud:2019rtt,Sirunyan:2018owy} 
in Fig.~\ref{fig:Higgs_DM}.
The comparison is performed in the context of Higgs portal models\cite{Patt:2006fw}. The translation of the 
H $\rightarrow$ invisible result
into a weak interacting massive particle–nucleon scattering cross section $\sigma_{\rm WIMP-N}$ relies on an effective
field theory approach under the assumption that invisible Higgs boson decays to a pair of WIMPs is
kinematically possible and that the WIMP is a scalar or a fermion\cite{Eboli:2000ze,Fox:2011pm,deSimone:2014pda}.
The excluded $\sigma_{\rm WIMP-N}$ values range down 
to e.g. 
$2 \times 10^{-46} {\rm cm}^2$ in the fermion WIMP
scenario, probing a new exclusion region for masses 
below 10 GeV.
When extrapolated to the full LHC dataset, including the high luminosity phase, a projected sensitivity of 2.5\%\ should be reached.

\section{The Higgs boson self-coupling}
\label{lambda}

In its Standard Model form of Eq.(\ref{eq:2b}), the self coupling $\lambda$ is related to the Higgs boson's mass and vev, as indicated in Eq.~(\ref{eq:2d}), and induces three Higgs and four Higgs boson interaction vertices after the spontaneous breaking of the electroweak symmetry. The measurement of the Higgs boson self interaction is of fundamental importance and the implications of such measurements are discussed in Section \ref{sec:vac}.

With knowledge of the Higgs boson mass, within the Standard Model the self-coupling of the Higgs boson became known. Its direct measurement is however crucial \cite{Carena:2018vpt}
to understand whether the electroweak symmetry breaking occurs as a crossover, as expected in the Standard Model, or as a strong first order phase transition in the early Universe, which would play a crucial role in our understanding of baryogenesis and has implications for possible gravitational wave signals. It is one of the deepest questions of the Standard Model and may provide a portal
to new physics beyond the Standard Model. 
Experimentally this can be addressed 
via the measurement of the production of multiple Higgs bosons.

With the run~2 dataset, a major effort to perform complete analyses in a number of final state channels, combining a variety of decay modes for each Higgs boson 
${\rm \gamma\gamma},$ ${\rm b\overline{b}}$, ${\rm \tau^+\tau^-}$ ,${\rm WW^*}$ seeking the highest sensitivity to the di-Higgs boson production was made~\cite{Aad:2019uzh,Sirunyan:2018ayu}. The final state that was immediately recognized for providing an optimal compromise between the number of events produced and a decent signal-to-background ratio, was ${\rm (H\rightarrow b\overline{b})(H\rightarrow \gamma\gamma)}$, but adding
additional channels can substantially enhance the 
sensitivity.
The net HH production 
receives contributions from processes that involve Higgs boson self-interactions as well as other interactions that are not sensitive to $\lambda$. The measurement of the kinematics of the HH system is crucial to disentangle the different contributions and infer information on $\lambda$~\cite{DiMicco:2019ngk}.
Individual experiment combinations with a partial run~2 dataset show that limits 
below 10
times the Standard Model expected rate can be set on the HH production cross section. An analysis with the full run~2 dataset from CMS in the ${\rm b\overline{b} \gamma\gamma}$ channel allows one to exclude values of the trilinear coupling 
smaller than -3.3
and larger than 8.5
times the Standard Model expectation~\cite{Sirunyan:2020xok}. With the full run~2 dataset, the ATLAS experiment performed a search for the ${\rm HH \rightarrow b \bar{b} b \bar{b}}$ final state in the electroweak vector boson fusion production mode~\cite{Aad:2020kub}, this process has a   cross section smaller by approximately one order of magnitude, but is sensitive to HHVV 
($c_{\rm 2V}$) coupling and has set a limit of $ -1.02 < c_{\rm 2V} < 2.71 $.
The sensitivity of these studies is still far from a measurement of the trilinear coupling.
However, it suggests
that with the full luminosity for the entire LHC program  a combined sensitivity of 4 standard deviations for observing HH production can be obtained~\cite{Cepeda:2019klc}. 
Another possibility is to constrain 
$\lambda$
indirectly through its loop level effect on single Higgs boson production
\cite{DiVita:2017eyz}.
However, such constraints are model dependent and typically rather weak.

\section{Searches for additional Higgs bosons
and non Standard Model interactions} 
\label{extrahiggses} 
 
While present data suggests the Higgs 
particle found is very Standard Model-like,
the experiments have tried nevertheless 
to crack and expose it as an
imposter, by searching for unexpected 
Higgs boson production processes or unexpected decays, but without success so far. Is there e.g. more than one Higgs or Higgs-like 
boson 
lurking in the wealth of data collected at the LHC? Within the 
 Standard Model only one fundamental 
 scalar is expected, but in many of its
 extensions the full Higgs family would contain several additional members. 
 Searches for both lighter and more massive (pseudo-)scalar particles, neutral and charged, have been 
 carried out, with no evidence found
 so far. A recent survey of the searches for additional Higgs bosons is reported in Ref.~\cite{Zyla:2020zbs}.

CP-violation is an essential ingredient required for our understanding of the matter anti-matter asymmetry in the Universe. Within the Standard Model very precise electric dipole moment measurements, in particular of the electron ~\cite{Andreev:2018ayy}
measured to be
$|d_e/e| < 1.1 \times 10^{-29}$~cm, impose very stringent limits on CP violation in the Higgs Yukawa sector~\cite{Brod:2013cka}. These constraints are model dependent, so probing directly the CP properties of the Higgs boson is 
mandatory.
The Standard Model Higgs boson is CP-even.
Non CP-even couplings of the Higgs boson
to vector bosons 
have been searched for in  Higgs boson 
decays to  ZZ$^*$ and WW$^*$~\cite{Aaboud:2017vzb,Sirunyan:2019twz} and through production in associated channels~\cite{Aad:2016nal,Sirunyan:2019nbs}
using several decay modes. 
CP-odd couplings
to fermions have  been searched for both in the decay of the Higgs boson to a pair of tau leptons~\cite{Berge:2011ij} and in the ${\rm t\overline{t}H}$ production mode with the subsequent decay of the Higgs boson to a pair of photons~\cite{Aad:2020ivc,Sirunyan:2019wxt}. No significant CP violating effects have been observed.

In the Standard Model the Yukawa couplings are diagonalized in the mass matrix; there are no off-diagonal terms. There are ways to evade this via effective dimension-6 operators yielding models with off-diagonal Yukawa couplings which are of  particular interest as these can break the relation between the masses and the couplings and can generate additional Higgs boson self-coupling terms. 
These operators describe
multi-particle correlations 
beyond the minimal Standard Model interactions
and are
suppressed by the square of some large mass scale which represents the scale of new physics.
The constraints from light quark or lepton flavour changing neutral currents, such as for instance $\mu \rightarrow e\gamma$, are very strong. However in the case of heavier fermions, e.g. from $\tau \rightarrow \mu \gamma$ or $\tau \rightarrow 3\mu$, these are  less stringent and the strongest bound on off-diagonal $\tau$-$\mu$ Yukawa couplings now comes from the search for Higgs boson decays to a tau-muon pair, with bounds on 
the mixing term
$|Y_{\tau\mu}|$ close to $10^{-3} \;$~\cite{Aad:2019ugc,Sirunyan:2017xzt}. 
LHC data 
so far revealed no evidence for higher dimensional operator correlations
divided by powers of any large mass scale
below the few TeV range
\cite{Slade:2019bjo,Ellis:2020unq}.

In the quark sector the existence of the Higgs boson at a mass of around 125~GeV has opened an interesting channel to search for flavour changing neutral current (FCNC) decays of the top quark to a Higgs boson and a charm or an up quark. FCNC top quark decays have been searched for in multiple subsequent Higgs boson decay channels, including the diphoton, ${\rm WW^*}$, $\tau^+ \tau^-$ and ${\rm b\overline{b}}$~\cite{HIGG-2013-09,TOPQ-2014-14,HIGG-2016-26,TOPQ-2017-15,TOPQ-2017-07,CMS-TOP-13-017,CMS-TOP-17-003}. Bounds on top FCNC decay branching fractions down to approximately 0.2\%\ are reached. These channels complement other FCNC top decays searches with a photon or a Z boson in the final state instead of the Higgs boson.

In summary, the Higgs
boson's properties 
measured at the LHC so far are consistent 
with the Standard Model expectations within
the present measurement precision. Furthermore, with a mass of 125~GeV, the effects of the predicted Higgs boson quantum corrections on electroweak observables within the Standard Model are entirely compatible with the precision measurements carried out at LEP and SLC, and at low energies~\cite{BAAK:2014gga}.

No evidence for additional 
Higgs bosons, Higgs boson decays into undetected particles, or CP, FCNC or lepton flavour violating effects in Higgs boson decays have been observed yet.  These tremendous successes of the Standard Model should however not keep us away from the fundamental open questions which remain unanswered. All these searches will therefore continue with the same vigour with anticipated much larger
 data sets in the future.

\section{Vacuum stability and hierarchies of scales}
\label{sec:vac}

If taken as the Standard Model Higgs boson,
the discovered boson 
completes Standard Model. 
It also comes with intriguing observations, including possible clues to 
new physics. 
How high in energy might the Standard Model work as our theory of particle interactions,
e.g. 
what is the ultraviolet limit of the Standard Model when taken as an effective theory?
What new interactions might lie beyond the Standard Model?
Important theoretical issues are the stability of the Higgs vacuum
and the small size of the electroweak scale
and Higgs boson's mass 
relative to the Planck scale,
$1.2 \times 10^{19}$ GeV,
where quantum gravity effects might apply.

\subsection{Vacuum stability}

\begin{figure}[t!]  
\centerline
{\includegraphics[width=0.60\textwidth]
{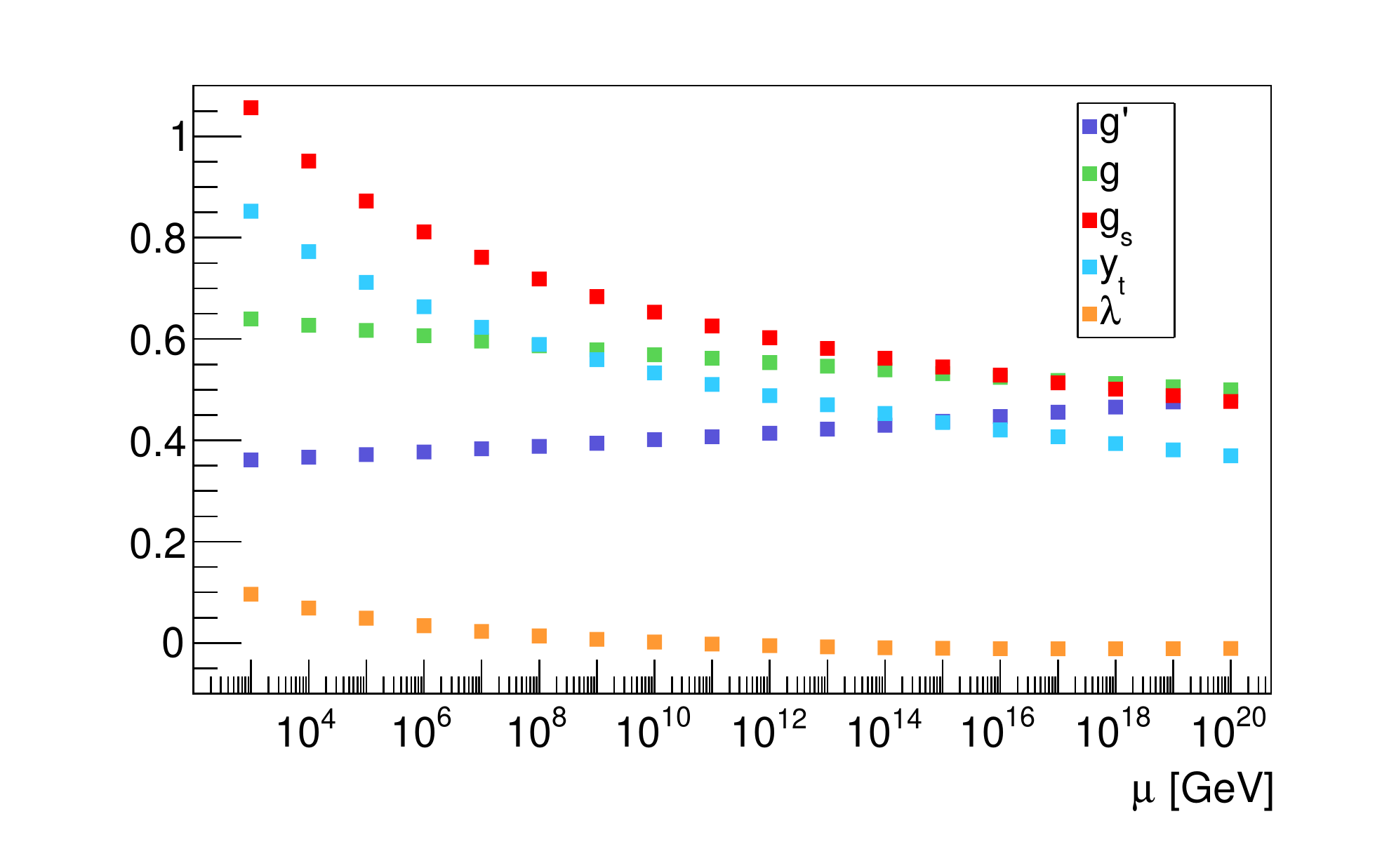}
\includegraphics[width=0.50\textwidth]
{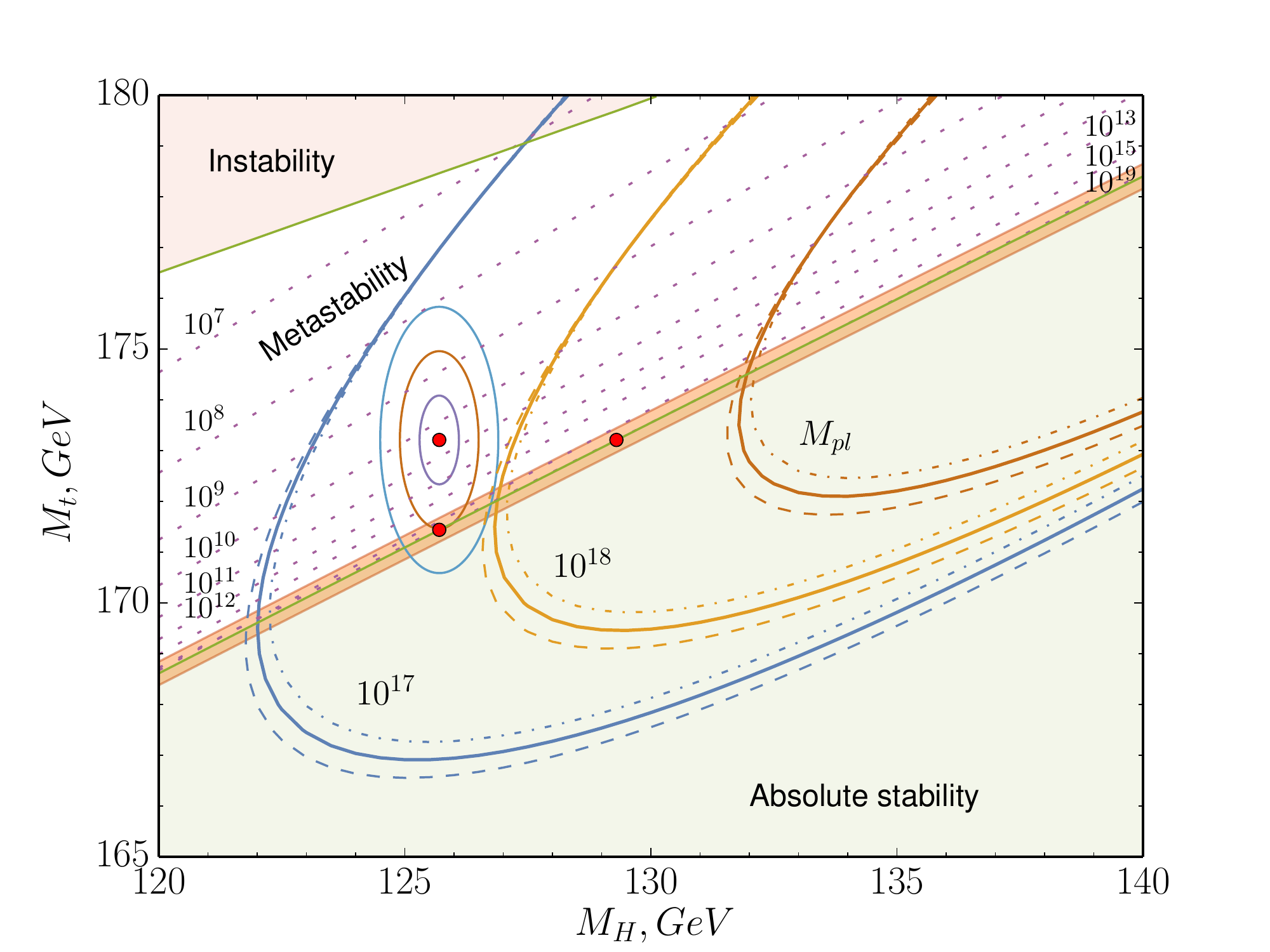}}
\caption{ 
Left: Running of the Standard Model
gauge couplings $g$, $g'$, $g_s$
for the electroweak SU(2) and U(1) and colour SU(3) interactions, 
the top quark Yukawa coupling $y_{\rm t}$ 
and Higgs boson self-coupling $\lambda$.
Figure from Ref.~\cite{Bass:2020nrg}
with couplings evaluated 
using the C++ code in Ref.~\cite{Kniehl:2016enc}.
Right:
Phase diagram of vacuum (meta)stability 
as function of the
top quark and Higgs boson masses
with one, two and three standard deviations ellipses.
The dotted lines refer to the scale where $\lambda$ touches zero. The parabola-like lines are contours describing the mass parameters leading to
vanishing rate of
change of $\lambda$
at fixed large scales.
Figure from Ref.~\cite{Bednyakov:2015sca}.
}%
\label{fig:couplings}
\end{figure}

With the discovered boson, 
the Standard Model is
perturbative and predictive when extrapolated to very high energies.
Interestingly,
if one extrapolates the 
Standard Model 
with 
its measured couplings up to the Planck scale 
and assumes no coupling to extra
   particles or new interactions, then the Higgs vacuum sits very close to the border of stable and metastable
   \cite{Degrassi:2012ry,Buttazzo:2013uya,Bezrukov:2012sa,Alekhin:2012py,Masina:2012tz,Hamada:2012bp,Jegerlehner:2013cta,Bednyakov:2015sca},
within 1.3 standard deviations of being stable 
\cite{Bednyakov:2015sca}.
With a metastable 
vacuum there would be a second minimum in the Higgs potential 
with value low than that measured at our energy scale (as shown in  Fig.~\ref{fig:higgs}
right)
after inclusion of radiative corrections.
For an unstable vacuum the BEH potential would become unbounded from below at large values of the BEH field.

Vacuum stability depends on the 
ultraviolet behaviour
of the 
Higgs boson self-coupling
$\lambda$.
The Standard Model
couplings
evolve with changing resolution 
(energy scale) according to the renormalization group 
as shown in the left panel of Fig.~\ref{fig:couplings}.
The weak SU(2) and QCD SU(3) couplings, $g$ and $g_s$ are
asymptotically free, with 
$\alpha_i = g_i^2/4 \pi$ 
decaying logarithmically 
with increasing resolution,
whereas the U(1) coupling $g'$
is non asymptotically free, rising in the ultraviolet. 
The top quark Yukawa coupling $y_{\rm t}$ decays with increasing resolution.
The running of the Higgs boson self-coupling $\lambda$ 
determines the stability of the electroweak vacuum.
Instability sets in if $\lambda$
crosses zero 
deep in the ultraviolet
and involves a delicate balance of Standard Model parameters.
With the Standard Model parameters measured at 
the LHC,
$\lambda$ decreases with 
increasing resolution.
This behaviour
is dominated by the large Higgs boson coupling to the top quark 
(and also QCD interactions of the top).
Without this coupling, $\lambda$ would rise in the ultraviolet.
In the left panel of 
Fig.~\ref{fig:couplings}
$\lambda$ crosses zero around $10^{10}$ GeV
with the top quark pole mass 
$m_{\rm t} = 173$ GeV and
$m_{\rm H}=125$ GeV.
This signals a metastable vacuum with lifetime 
greater than about $10^{600}$ years~\cite{Buttazzo:2013uya}, very much greater than the present age of the Universe, about 13.8 billion years -- see the right panel in Fig.~\ref{fig:higgs}.
The right panel of 
Fig.~\ref{fig:couplings}
indicates the sensitivity of vacuum stability to small changes in $m_{\rm t}$.
If the top mass is
 taken as 171 GeV in these calculations,
the vacuum stays stable up to the Planck scale.
The measured 125 GeV Higgs boson mass is close to the minimum needed
for vacuum stability with the measured top quark mass.

The Standard Model observed
in our experiments, assuming no extra particles at higher energies, 
is strongly correlated with its behaviour in the extreme ultraviolet.
Might 
this be telling us something deep about the origin of the Standard Model?

It is important to emphasise the large extrapolations in these calculations.
New physics even at the largest scales can change this picture~\cite{Branchina:2013jra}. 
Modulo this caveat,
the Higgs vacuum sitting ``close to the edge'' of stable and metastable suggests possible 
new critical phenomena in the deep ultraviolet.
One possible
interpretation is a statistical system in 
the ultraviolet, near to the Planck scale, 
close to its
critical point
\cite{Jegerlehner:2013cta,Degrassi:2012ry,Buttazzo:2013uya}.
As a general rule, theories with new interactions and new particles at higher energies 
should make the vacuum more stable rather than less!

\subsection{
Scale hierarchies and
the origin of Standard Model gauge symmetries}

The Higgs boson's mass is very much less than
the Planck scale despite quantum corrections which naively act to push its mass towards the deep ultraviolet.
Under renormalization, the Higgs boson's mass squared comes with a quadratically divergent counterterm
which comes from the Higgs boson self-energy, viz. 
\begin{equation}
m_{{\rm H \ bare}}^2 
= m_{{\rm H \ ren}}^2 + \delta m_{\rm H}^2,
\label{eq:4a}
\end{equation}
where
\begin{equation}
\delta m_{\rm H}^2 
=
\frac{K^2}{16 \pi^2}
\frac{6}{v^2} 
\biggl(
m_{\rm H}^2 + m_{\rm Z}^2 + 2 m_{\rm W}^2 - 4 m_{\rm t}^2
\biggr)
\label{eq:4b}
\end{equation}
relates the renormalized and bare Higgs boson 
masses and
we neglect small contributions from lighter mass quarks.
Here 
$K$ is an ultraviolet cut-off scale on the momentum integrals characterizing the limit to which the Standard Model should work.
If $K$ is taken as a physical scale, e.g., the
Planck scale,
then why is the physical Higgs boson's mass so small compared to the cut-off?
This topic, 
called the hierarchy or
naturalness puzzle,
has attracted much theoretical attention~\cite{Giudice:2008bi,Wells:2009kq}.
What stabilizes the value of $m_{\rm H}$? 
One possibility is that the Higgs boson's mass is fine tuned, perhaps through some kind of environment selection and perhaps in connection with vacuum stability of the Standard Model.
Alternatively, the Standard Model
quantum correction to the Higgs boson's mass, 
which is dominated 
by the top quark contribution,
might be cancelled by any new particles that couple to the Higgs boson.
However, such particles have so far not been seen 
in the mass range of the LHC.
Likewise, any composite structure to the Higgs boson would soften the ultraviolet divergences
but there is no evidence for this in the present data.
Searches for extra particles
and possible composite structure will continue in the next years with increased luminosity at the LHC.

The Standard Model is a mathematically consistent theory 
up to the Planck scale but is it also physically complete up to the Planck scale?
Some new physics is needed
but the scale where it first appears is an open issue.
It would be very surprising if there is no new physics below the Planck scale. 
In the absence of new physics one has the naturalness puzzle, 
which has
inspired much thinking about possible extra particles.
Theoretical attempts to resolve this include weakly coupled models
with a popular candidate being supersymmetry~\cite{Wess:1973kz}, which if present in Nature would be a new symmetry between bosons and fermions.
Strongly coupled models where the Higgs boson is considered as a bound state of new dynamics strong at the weak scale are an alternative solution. 
Here the "lightness" of the Higgs boson can be explained if the Higgs boson turns out to be a pseudo-Nambu–Goldstone boson. Such models include 
the so called 
little Higgs~\cite{ArkaniHamed:2001nc,ArkaniHamed:2002qy}, 
twin Higgs~\cite{Chacko:2005pe}
and partial compositeness~\cite{Kaplan:1991dc}
models. 
For a comprehensive review of these ideas and
their phenomenology
see 
Chapter 11 of 
Ref.~\cite{Zyla:2020zbs},
with possible alternatives to an elementary Higgs boson also discussed in Ref.~\cite{Csaki:2015hcd}. 
A related issue is the deeper origin of the gauge symmetries of particle physics.
Might the electroweak and QCD interactions unify in the ultraviolet within some larger gauge group?
The ultimate dream of this approach is unification of the Standard Model forces 
with gravity 
close to the Planck scale.
With unification one expects the 
running gauge couplings 
of the Standard Model
to meet in the ultraviolet. 
They do come close -- see Fig.~\ref{fig:couplings} 
--  
but without exact crossing. 
This could be achieved with the addition of 
supersymmetry, SUSY, at TeV energies~\cite{Ross:1992tz},
which might also provide a dark matter candidate particle.
While the simplest SUSY 
models would
have liked a Higgs boson mass close to measured value, the absence of any signal for new SUSY particles in LHC experiments
means that these models are now strongly constrained
\cite{Altarelli:2013lla,Altarelli:2014xxa,Pokorski:2016dne,Ross:2014mua}.
The present status of minimal SUSY model predictions for Higgs boson mass(es) is discussed in Ref.~\cite{Slavich:2020zjv}.
Any new symmetries in the ultraviolet must be strongly broken so they are not seen at the energies of our experiments
meaning that there is a trade off: the extra symmetry that might exist at higher energies also comes with a (perhaps large) number of new parameters 
needing extra explanation.

Modulo the large extrapolations involved and any new particles waiting to be discovered at higher energies, it is interesting not to  discard the idea that the Standard Model might work to very high energies close to the Planck scale.
In this alternative scenario, it is plausible that the Standard Model might 
behave as an emergent effective theory
with gauge symmetries
``dissolving'' in the extreme ultraviolet
\cite{Jegerlehner:2013cta,Jegerlehner:1998kt,Bjorken:2001pe,Forster:1980dg,Giudice:2017pzm,Witten:2017hdv,Bass:2020gpp}.
That is, the Standard Model particles 
including the Higgs and gauge bosons
could be the long-range, collective excitations of a statistical system near to its critical point
that resides close to 
the Planck scale~\cite{Jegerlehner:2013cta}.
Emergent gauge symmetries, where we make symmetry as well as breaking it, are important in 
quantum many body systems
such as high temperature superconductors~\cite{Baskaran:1987my}, 
in topological phases of matter \cite{Sachdev:2018ddg}
and 
in the low energy limit of the Hubbard model~\cite{Affleck:1988zz} employed in quantum simulations of gauge theories \cite{Banerjee:2012pg,Banuls:2019bmf}.
Emergent gauge symmetry can arise associated with an infrared fixed point in the renormalization group
\cite{Wetterich:2016qee}.

Gauge symmetry and
renormalizability constrain the global symmetries of the
Standard Model
at mass dimension four~\cite{Witten:2017hdv}.
If the Standard Model is an
effective theory emerging in the infrared,
low-energy global symmetries such as lepton and baryon-number conservation can 
be broken through additional 
(non-renormalizable) higher dimensional terms,
suppressed by powers of a large mass scale $M$ that characterizes the ultraviolet limit of the effective theory~\cite{Witten:2017hdv,Bass:2020gpp,Weinberg:2018apv,Jegerlehner:2013cta}. 
The tiny neutrino masses 
suggested by neutrino oscillation data
\cite{Balantekin:2018azf} have a simple interpretation in this picture.
If neutrinos with zero electric charge
are Majorana particles,
meaning they are their own antiparticles,
then their masses can be linked to
lepton number violation and
the dimension five Weinberg operator 
\cite{Weinberg:1979sa},
suppressed by single power of $M$,
involving just the lepton and Higgs fields with
$m_{\nu} \sim \Lambda_{\rm ew}^2/M$
where
$\Lambda_{\rm ew} = 246$ GeV
is the electroweak scale
and
$M \sim 10^{15}$ GeV. 
If, instead, neutrinos are Dirac particles with their tiny masses coming from Yukawa couplings to the Higgs field, one 
then has to ask
why these Yukawa couplings are so much suppressed relative to the charged lepton couplings with right-handed neutrinos not participating in electroweak interactions.

\section{The Higgs boson and cosmology}
\label{sec:cosmo}

The Higgs boson influences many ideas in cosmology, from thinking about the accelerating expansion of the Universe today to processes in the early Universe.

The Higgs potential generates 
a large vacuum energy contribution to the cosmological constant
or vacuum energy density 
$\rho_{\rm vac}$,
a prime candidate for the dark energy that drives the accelerating expansion of the Universe.
Astrophysics experiments 
\cite{Aghanim:2018eyx}
tell us
that
$\rho_{\rm vac} = (0.002 \ {\rm eV})^4$.
In the Standard Model
$\rho_{\rm vac}$
receives contributions from the Higgs and QCD condensates, the zero-point energies of quantum field theory 
and also a gravitational contribution
\cite{Weinberg:1988cp}.
Electroweak and QCD contributions are characterized by scales of 246 GeV and 200 MeV, so what cancels them to give the net cosmological constant scale 0.002 eV?
This puzzle has attracted considerable theoretical attention and ideas, see e.g. Refs.~\cite{Weinberg:1988cp,Wetterich:1994bg,Veltman:1997nm,Sahni:1999gb,Peebles:2002gy,Copeland:2006wr,Straumann:2007zz,Frieman:2008sn,Bass:2011zz,Martin:2012bt,Dvali:2014gua}.
Present measurements are consistent with a time independent cosmological constant.
The next generation of cosmology surveys will look for any time dependence of dark energy, 
with sensitivity to
any variations from a time independent
cosmological constant of 10\% or 
more~\cite{laureijs2011euclid}.

An intriguing issue is the similar size of the cosmological constant 
scale 0.002 eV 
to what we expect for 
the value of
light neutrino masses 
\cite{Altarelli:2004cp}.
With a finite cosmological constant, there is no solution of Einstein's equations of General 
Relativity with constant Minkowski metric $g_{\mu \nu}$. 
Global space-time  translational invariance 
of the vacuum is broken by a finite cosmological constant~\cite{Weinberg:1988cp}.
Motivated by the success of the Standard Model
and Special Relativity
in our experiments, 
one might 
suppose that the vacuum including condensates with finite vevs
is space-time translational invariant 
and 
that flat space-time is consistent at dimension four.
Then, if 
the Standard Model is treated as an 
effective theory emergent below 
a large ultraviolet scale $M$, 
the 
global symmetry might be broken through higher
dimensional terms
with the 
electroweak and QCD scales
$\Lambda_{\rm ew}$ and 
$\Lambda_{\rm qcd}$
entering the
cosmological constant
with the scale of the leading term
suppressed by 
$\Lambda_{\rm ew}/M$
(that is, with vacuum energy density $\rho_{\rm vac} \sim (\Lambda_{\rm ew}^2/M)^4$
 with one factor of 
 $\Lambda_{\rm ew}^2/M$ for each dimension of space-time)
 \cite{Bass:2020nrg,Bass:2020egf}. 
How the Higgs potential relates to space-time
structure is an important issue for cosmology.

One of the main ideas for understanding the matter-antimatter asymmetry in the Universe involves a
possible 
first order phase transition with the Higgs vev generated in the early Universe
\cite{Trodden:1998ym,Morrissey:2012db}. Bubbles with Higgs condensate would be created and expand at the speed of light. This scenario also needs 
new sources of CP violation beyond the usual Standard Model, with recent ideas discussed in
\cite{Servant:2018xcs}
together with an 
extended Higgs sector, e.g. extra singlet scalar, 
and 
quantum tunneling 
processes in the vacuum 
called sphalerons that violate baryon number.
Evidence of any
first order electroweak phase transition might 
show up in future gravitational wave measurements with
LISA, the Laser Interferometer Space Antenna mission of the European Space Agency ESA \cite{Audley:2017drz,Caprini:2015zlo} and the proposed AEDGE experiment \cite{Bertoldi:2019tck}.

Besides accelerating expansion today,
the Universe is commonly believed to have undergone an initial period of exponential
expansion called inflation, with factor at least $10^{26}$ in the first
about $10^{-33}$ seconds \cite{Baumann:2008bn}.
Inflation is posited to explain the uniformity with small anisotropies in the Cosmic Microwave Background, CMB.
The particle physics of inflation is 
unknown, though there are many theoretical ideas
including where the Higgs boson plays an important 
role~\cite{Bezrukov:2007ep,Jegerlehner:2014mua,Rubio:2018ogq}
or 
where the vev of a possible time dependent extra scalar field
interpolates between initial inflation and dark energy today~\cite{Wetterich:1994bg,Wetterich:1987fm,Peebles:1987ek,Peebles:2002gy}. 
Possible extra fundamental scalars also enter 
in extended theories of gravitation, beyond minimal General Relativity~\cite{Capozziello:2011et}.
One idea of inflation 
connects the Higgs boson to gravitation with non-minimal coupling
$\xi$ H$^2$R
where R is the Ricci curvature scalar of General Relativity~\cite{Bezrukov:2007ep}.
The parameters of this model can be chosen to reproduce key features of the CMB,
the measured scalar spectral index and the tensor-to-scalar ratio.

The Higgs boson may thus play an essential role in the evolution of the Universe from its very beginning to the physics we see in our experiments today!

\section{Conclusions: Outlook and future measurements - 
towards a future Higgs 
factory}
\label{conclusions}

The Higgs boson discovered at the LHC is the first ever
observed scalar elementary particle, and we are just at the beginning in understanding what this
particle can teach us.
Are there maybe other (pseudo)scalars out there? What is their role?

The precision reached on Higgs boson properties so far had not been anticipated. Channels have been measured which were thought to be inaccessible. This extraordinary success is due to several factors: (i) the remarkable operation of the LHC; (ii) the robustness of the experiment; (iii) the improvements in experimental techniques; (iv) the revolutions in the theoretical prediction and simulation of the LHC processes; and (v) the {\it "gift of Nature"} that the mass of the Higgs boson be precisely at the cross roads of all possible decay modes.

The found Higgs boson  so far behaves very Standard Model like
with 
its couplings
to the W and Z gauge bosons, 
to the third family of charged fermions
and to muons
from the second family, all
consistent with the BEH mechanism of mass generation. But presently we do not yet know 
why the particle masses have the values
as observed in Nature.
Furthermore, if the Standard Model is extrapolated up to the Planck scale
with the measured Higgs boson and top quark parameters, then the
electroweak vacuum sits very close to the border of the stable and metastable regions.

\begin{figure}[t!]  
\centerline
{\includegraphics[width=0.67\textwidth]
{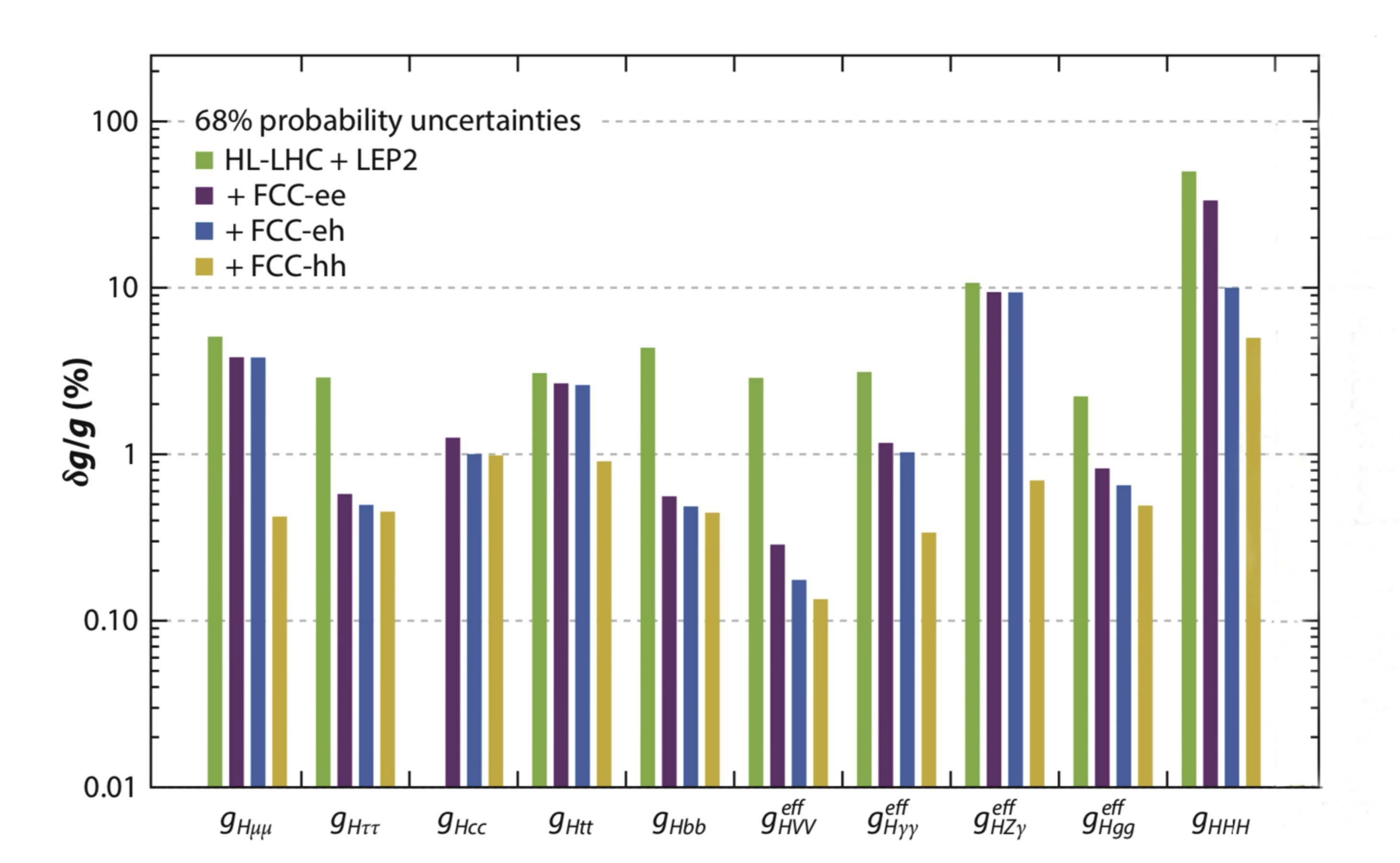}
}
\caption{
1$\sigma$ precision reach at the FCC on the effective Higgs boson couplings to fermions (muons, taus, charm quarks, top quarks and b quarks), to vector bosons ($W$ or $Z$, photons, $Z\gamma$ and gluons) and the Higgs boson self-coupling 
in an effective field theory framework. Absolute precision in the electroweak measurements is assumed. The different bars illustrate the improvements that would be possible by combining each FCC stage with the previous knowledge at that time.
Abbreviations: FCC, 
Future Circular Collider; FCC-ee, FCC electron–positron collider; FCC-eh, FCC electron–proton collider; FCC-hh, FCC hadron collider; HL-LHC, High-Luminosity Large Hadron Collider; LEP2, Large Electron–Positron Collider.
Figure taken from \cite{Abada:2019lih}. %
}
\label{fig:future}
\end{figure}

Measurements will be continued with the LHC run~3 starting in 2022, doubling the collision statistics recorded till now, 
and
later with the high luminosity upgrade of the LHC \cite{Bruning:2019abc}
starting before the end of this decade, 
for a factor 20 increase in the total statistics.
These measurements will enable considerable increase in precision on our knowledge of the Higgs boson's couplings
to a precision of a few \% for many of these, 
as well to make first measurements of the Higgs boson's self-interaction, as shown in Fig.~\ref{fig:future}.
Along with increased  experimental precision, there is also need for 
more precise theoretical technology~\cite{Heinrich:2020ybq}
for an optimal 
extraction of key quantities from the data. 
Scientists at the LHC secretly hope that the 
precision measurements will show sooner or later deviations from these predictions, and thus the first cracks in the Standard Model.

The Standard Model is an extraordinary theory, fully consistent up to very high energy scales. We know however that it should not be the whole story. 
How high in energy will it continue to hold
before new physics sets in? 
Most likely,
the Higgs boson
will provide us with an
essential
window to probe this new world beyond the Standard Model.

Beyond the Standard Model searches 
include looking for
any CP violating couplings of the Higgs boson, any extended Higgs sector with possible connections to baryogensis, 
possible composite structure to the Higgs boson, 
which could provide indications of unknown underlying dynamics which could be responsible for the BEH mechanism,
as well as any decays to possible dark matter candidates which might
explain the mysterious 80\% missing mass component in the Universe.

The recent European Particle Physics Strategy 
update \cite{Gianotti:2020abc,EuropeanStrategyGroup:2020pow}
highlights precision studies of the Higgs boson and its interactions as the main priority for the next high-energy collider, 
with long term options post the LHC luminosity upgrade including a Future Circular Collider 
\cite{Benedikt:2019abc,Benedikt:2020ejr} (FCC-ee)
or CLIC linear $e^+ e^-$ collider  \cite{Stapnes:2019abc,Sicking:2020gjp}
being considered in context of the future of CERN, with additional
$e^+ e^-$ collider options being discussed such
as the ILC
linear collider in Japan
\cite{Michizono:2019abc} and a circular collider CEPC in China~\cite{Lou:2019abc}. The circular 
collider projects also include a 
proton-proton (FCC-hh) and
proton-lepton (FCC-eh) option, typically planned as 
a next stage following the $e^+e^-$ option.
A detailed discussion of the precision that can be reached on Higgs boson property measurements with
 these different options is reported in
 \cite{deBlas:2019rxi}, with an example for the 
 FCC facility shown in Fig.~\ref{fig:future}.
 Electron-positron colliders cannot reach as 
 high centre of mass energies as proton-proton 
 colliders, but benefit from a much cleaner 
 initial collision state consisting of fundamental particles with well defined energies entering the
 interactions. In particular Higgs production will
 be measured inclusively from its presence as a 
 recoil to the Z in $e^+e^- \rightarrow {\rm HZ}$ events,
 allowing the absolute measurement of the 
 Higgs boson's coupling to the Z boson.

 As shown in Fig.~\ref{fig:future} 
 sub-percent precisions on the couplings of the Higgs boson to other gauge bosons and charged fermions of second and third generation, 
 except the strange-quark, can  be achieved in $e^+e^-$ collisions where the precision coupling of the Higgs boson to gauge bosons will reach the per
 mille level, the couplings to bottom-quarks, taus and muons will reach the level of 4 per mille and the coupling to charm-quarks the percent level~\cite{deBlas:2019rxi}. 
 In addition, the prospect of measuring, or at least strongly constraining, the couplings to the three lightest quarks and to the electron by dedicated FCC $e^+e^-$ runs at the 
 Higgs boson's mass, are being evaluated.
 
 The direct measurement of the Yukawa coupling to top quarks requires high-energy proton-proton collisions or higher centre-of-mass energies in $e^+e^-$ and should reach the percent level precision.
 The study of invisible Higgs boson decays will reach sensitivities well below the percent level and reach the level of the expected 
 Standard Model rate, 
driven by Higgs bosons decaying in to a pair of Z bosons, which each decay into neutrinos.

To improve significantly on the precision of Higgs boson
self-coupling with direct measurements will require high-energy hadron beams, in lepton-hadron or 
hadron-hadron mode operation,
with the highest expected precision of approximately 5\%\ being obtained at a very high-energy proton-proton collider with centre-of-mass energy of 100~TeV~\cite{deBlas:2019rxi}.

Besides the charged leptons and quarks,
the Higgs boson might also play a key role in generation of neutrino masses, where if neutrinos are their own antiparticles the neutrino mass enters through the 
dimension-five Weinberg operator.
Neutrinoless double $\beta$ decay experiments aim to look for Majorana neutrinos
with the next generation experiments
sensitive to the
theoretically interesting mass range
\cite{Agostini:2017jim,Caldwell:2017mqu}.
The role of the Higgs potential in the vacuum energy density of the Universe is important to understanding the dark energy that drives the accelerating expansion of the Universe. Signals of possible phase transitions in the early Universe, e.g., responsible for baryogenesis, might be manifest in future gravitational wave measurements.

This rich program of experiment and theory promises to shed exciting new insights into the Higgs boson
and its vital role in the physics of the Universe!

\newpage

\bibliography{sample}

\begin{thebibliography}{100}
\bibitem{Foot1}
\bibinfo{note}{The electronvolt, eV, is an energy unit used in particle physics
  corresponding to about $1.6 \times 10^{-19}$ Joules, with 1 TeV = $10^{12}$
  eV.}
\bibitem{Foot2}
\bibinfo{note}{A fb$^{-1}$ is a unit that quantifies the integrated luminosity.
  It has the dimension of inverse area, proportional to the amount of
  proton-proton collisions produced by the LHC collider. One femtobarn 1fb $=
  10^{-43}$m$^2$. One inverse attobarn 1ab$^{-1}$ = $10^{3}$fb$^{-1}$. One
  inverse femtobarn corresponds to approximately $10^{14}$ proton-proton
  collisions in the LHC.}
\end{thebibliography}

\section*{Acknowledgements}
None of the results presented in this review 
would have been possible without the 
diligent efforts of all our colleagues from the LHC accelerator group, the ATLAS and CMS experiments, the computing divisions, the theoretical community
and many more, 
whom all took part in this fantastic adventure at the energy frontier. Specifically we like to thank Maria Cepeda, Fred Jegerlehner and
Janina Krzysiak
for sharing
their insight and useful 
discussions in the preparation of this manuscript.

\end{document}